\documentclass[nohyperref]{article}

\usepackage{microtype}
\usepackage{graphicx}
\usepackage{subfigure}
\usepackage{booktabs} 

\usepackage{hyperref}

\usepackage[accepted]{icml2023}

\usepackage{amsmath}
\usepackage{amssymb}
\usepackage{mathtools}
\usepackage{amsthm}
\usepackage{physics}
\usepackage{cuted}
\usepackage{listings}

\usepackage[capitalize,noabbrev]{cleveref}

\theoremstyle{plain}
\newtheorem{theorem}{Theorem}[section]

\theoremstyle{definition}

\theoremstyle{remark}

\newcommand{\es}{\epsilon}

\usepackage[textsize=tiny]{todonotes}

\icmltitlerunning{Q-Flow: Generative Modeling for Differential Equations of Open Quantum Dynamics with Normalizing Flows}

\begin{document}

\twocolumn[
\icmltitle{Q-Flow: Generative Modeling for Differential Equations of Open Quantum Dynamics with Normalizing Flows}




\begin{icmlauthorlist}
\icmlauthor{Owen Dugan}{mitp,iaifi}
\icmlauthor{Peter Y. Lu}{chi,iaifi}
\icmlauthor{Rumen Dangovski}{mite,iaifi}
\icmlauthor{Di Luo}{iaifi,ctp,har}
\icmlauthor{Marin Solja\v{c}i\'{c}}{mitp,iaifi}
\end{icmlauthorlist}

\icmlaffiliation{mitp}{Department of Physics, Massachusetts Institute of Technology}
\icmlaffiliation{mite}{MIT EECS}
\icmlaffiliation{chi}{Data Science Institute, University of Chicago}
\icmlaffiliation{iaifi}{NSF AI Institute for Artificial Intelligence and Fundamental Interactions}
\icmlaffiliation{ctp}{Center for Theoretical Physics, Massachusetts Institute of Technology}
\icmlaffiliation{har}{Department of Physics, Harvard University}

\icmlcorrespondingauthor{Di Luo}{diluo@mit.edu}

\icmlkeywords{Machine Learning, ICML}

\vskip 0.3in
]



\printAffiliationsAndNotice{}  

\begin{abstract}
Studying the dynamics of open quantum systems can enable breakthroughs both in fundamental physics and applications to quantum engineering and quantum computation. Since the density matrix $\rho$, which is the fundamental description for the dynamics of such systems, is high-dimensional, customized deep generative neural networks have been instrumental in modeling $\rho$. However, the complex-valued nature and normalization constraints of $\rho$, as well as its complicated dynamics, prohibit a seamless connection between open quantum systems and the recent advances in deep generative modeling. Here we lift that limitation by utilizing a reformulation of open quantum system dynamics to a partial differential equation (PDE) for a corresponding probability distribution $Q$, the Husimi Q function. 
Thus, we model the Q function seamlessly with \emph{off-the-shelf} deep generative models such as normalizing flows. Additionally, we develop novel methods for learning normalizing flow evolution governed by high-dimensional PDEs based on the Euler method and the application of the time-dependent variational principle. We name the resulting approach \emph{Q-Flow} and demonstrate the scalability and efficiency of Q-Flow on open quantum system simulations, including the dissipative harmonic oscillator and the dissipative bosonic model. Q-Flow is superior to conventional PDE solvers and state-of-the-art physics-informed neural network solvers, especially in high-dimensional systems.
\end{abstract}

\section{Introduction}\label{sec: intro}

Understanding open quantum system dynamics is crucial for fundamental physics and high-impact scientific applications such as quantum engineering and quantum computation~\citep{verstraete2009quantum,barreiro2011open}. 

\begin{figure}[t]
    \centering
    
    \includegraphics[width=\linewidth]{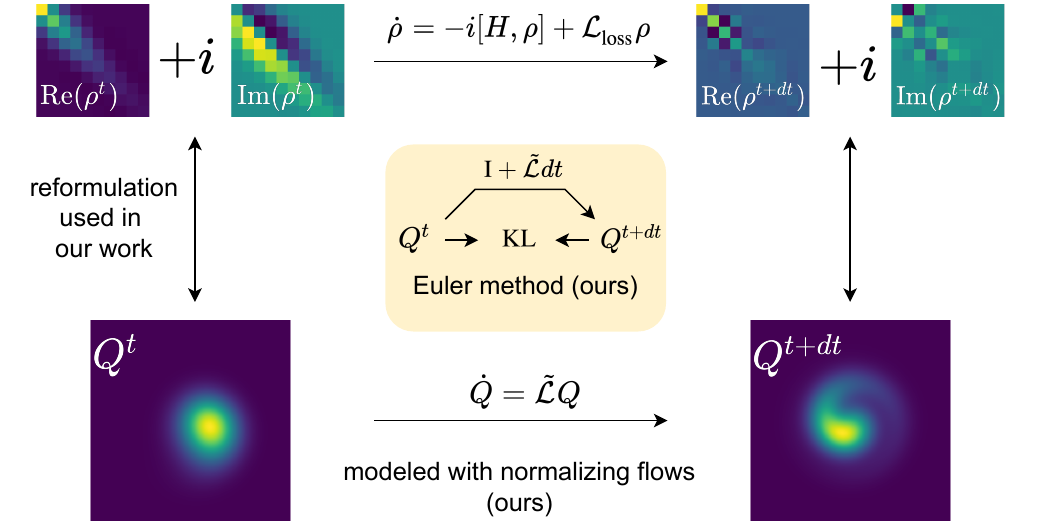}
    
    \caption{Q-Flow. We reformulate differential equations for density matrix dynamics as PDEs for probability distribution dynamics. We use off-the-shelf normalizing flows and our Euler-KL method for solving such PDEs.}
    
    \label{fig:main}
\end{figure}

The state of an open quantum system is given by the density matrix $\rho$, which is an exponentially scaling object with a size that grows as $N^{2k}$ for $k$ subsystems each with a Hilbert space of dimension $N$. Thus, computing or evolving $\rho$ becomes infeasible as $k$ increases due to the curse of dimensionality. 
Pioneering work on representing $\rho$ in a compact form as a customized deep generative neural network has shown great promise in advancing the frontier of understanding high-dimensional quantum systems~\citep{PhysRevLett.122.250503,PhysRevB.99.214306,PhysRevLett.122.250502,PhysRevLett.122.250501}. However, a number of computational challenges remain when solving for $\rho$, which motivates the development of novel machine learning methods. Notable challenges are:
\begin{enumerate}
    \item The density matrix $\rho$ is \emph{complex-valued} and has the constraint $\tr[\rho]=1$. That makes it non-trivial to model with standard generative models that are real-valued.
    \item The differential equation that governs the dynamics of $\rho$ models complicated interactions in high-dimensional space, which hampers the application of conventional differential equation solvers.
    \item Previous efforts to model $\rho$ with neural networks are restricted to discrete spin systems, and it is unclear how to model $\rho$ in continuous or bosonic systems.
\end{enumerate}

The state-of-the-art literature has addressed Challenge 1 by developing customized neural architectures for $\rho$ in spin systems with discrete degrees of freedom only~\citep{PhysRevLett.122.250503,PhysRevB.99.214306,PhysRevLett.122.250502,PhysRevLett.122.250501,luo2022autoregressive,reh2021time}. Challenge 2 has been attempted by exploring physics-inspired training objectives, such as Physics-informed neural networks (PINNs)~\citep{JMLR:v19:18-046,RAISSI2019686}. PINNs have shown promise in low-dimensional systems, but it is not clear how they can be scaled to high-dimensional PDEs. Furthermore, the existing literature has not addressed Challenge 3 and missed an opportunity to establish a \emph{direct} connection between modeling continuous variable open quantum dynamics and novel deep generative models for standard machine learning benchmarks. Such a connection would drive progress in both machine learning applications for open quantum dynamics and deep generative modeling.

In this paper, we address Challenges 1-3 by establishing a bridge between open quantum systems in a continuous Hilbert space and continuous-variable generative modeling.

Firstly, we reformulate the problem by replacing the density matrix $\rho$ with an alternative representation, the Husimi Q function $Q$~\citep{QuantumOpticsChapter4}, which can be practically considered as a probability distribution. Thus, we enable the use of \emph{off-the-shelf} generative neural networks to model $Q$. Because the optimization of high dimensional quantum systems requires access to both easy sampling and probability density values, we use \emph{normalizing flows}~\citep{dinh2014nice, rezende2015variational} as our generative model.

Secondly, we develop novel methods for training normalizing flows that obey complicated high-dimensional PDEs, which are an excellent fit for approximating $Q$. We propose a training method, the stochastic \emph{Euler-KL} method, which is based on the forward discretization of the differential equation for $Q$ and the Kullback-Lieber matching of probability distributions. Our normalizing flows approach can also be equipped with the Time-Dependent Variational Principle (TDVP) method~\citep{mcmillan1965ground}, which can be derived from the Euler method and can be thought of as an analog of the natural gradient method~\citep{amari1996neural,amari1998natural}.

We name our contributions \emph{Q-Flow} (see Figure~\ref{fig:main}). Q-Flow is a new approach to solving open quantum systems based on off-the-shelf normalizing flows and the Euler/TDVP methods for evolving such flows in complicated PDEs. We demonstrate that Q-Flow is scalable and efficient for simulating various open quantum systems. Our contributions can be summarized as follows:
\begin{itemize}
\item A new generative modeling approach for open quantum dynamics with continuous degrees of freedom based on the Husimi Q function, which allows for using normalizing flows off the shelf.
\item New methods for solving open quantum dynamics PDEs using normalizing flows with stochastic Euler-KL method and TDVP.
\item Demonstration of the scalability and efficiency of our methods on simulations of dissipative harmonic oscillator and dissipative bosonic models by surpassing conventional PDE solvers and state-of-the-art machine learning PDE solvers, physics-informed neural networks (PINN).
\end{itemize}

Importantly, with Q-Flow, the difficulty in simulating quantum dynamics is no longer the dimension of the simulation but instead the complexity of the Q function and its evolution, which opens a new avenue for research.

\section{Related Work}

\subsection{Neural Network Quantum States}

Neural network quantum states are generative neural network architectures---including restricted Boltzmann machines \cite{doi:10.1126/science.aag2302}, autoregressive models \cite{PhysRevLett.124.020503,PhysRevLett.128.090501,chen2022simulating,luo2021gauge}, and determinant neural network models \cite{PhysRevResearch.2.033429,Hermann2020,luo2019backflow}---that have been adapted to represent quantum wave functions or density matrices (in the case of open quantum systems) rather than probability distributions. They are optimized using variational quantum Monte Carlo methods and have primarily been applied to model discrete spin systems \cite{doi:10.1126/science.aag2302,PhysRevLett.124.020503,PhysRevLett.128.090501} as well as tackle the continuous many-body wave function in quantum chemistry applications~\cite{PhysRevResearch.2.033429,Hermann2020} and quantum field theories~\cite{luo2022gauge,martyn2022variational}.

In contrast with prior deep learning-based approaches that directly model the wave function or density matrix, our work focuses on the Q function representation of the quantum state---a continuous quasiprobability distribution \cite{QuantumOpticsChapter4} that can be modeled using an appropriate generative model, e.g.,\ normalizing flows.

\subsection{Partial Differential Equation (PDE) Solvers}

To model the dynamics of an open quantum system using the Q function formulation, we are required to solve a high-dimensional PDE. By parameterizing the Q function using a normalizing flow, our approach can efficiently solve this PDE. For comparison, we benchmark our work against alternative PDE solvers.

Traditional PDE solvers struggle to handle high-dimensional PDEs due to the curse of dimensionality, where storing the state of the system on a grid or mesh grows expontentially with the dimension of the problem. As traditional solver benchmarks, we use finite-difference and pseudo-spectral methods~\citep{courant1928uber,fornberg1998practical}. While there are specialized methods for solving high-dimensional PDEs, they are often complex to set up and only apply to a few restricted classes of PDEs, e.g.,\ parabolic PDEs \cite{weinan2021algorithms}. From such specialized methods, we benchmark against a Stochastic method \cite{martin_reining_ceperley_2016}.

We also benchmark against physics-informed neural networks (PINNs)---a promising deep learning-based variational approach for solving PDEs \cite{JMLR:v19:18-046,RAISSI2019686,BERG2019239}. PINNs, however, have been shown to have limitations related to the difficulty of the variational optimization problem \cite{NEURIPS2021_df438e52} and, in their standard form, may also suffer from the curse of dimensionality.

\section{Solving Open Quantum Dynamics with Q-Flow}

In this work, we develop Q-Flow, an approach to solving open quantum dynamics based on flow-based models under the Q function partial differential equation formulation. The key contributions of our work are twofold. Firstly, we establish a general framework for solving open quantum dynamics learning through the flow-based model representation. Secondly, we develop optimization algorithms for solving high dimensional partial differential equations and apply them to PDEs for the Q function. Note that a more thorough review of the relevant Quantum Mechanics is provided in Appendix \ref{sec:quantum_prelim}.

\subsection{Quantum Overview}

The fundamental mathematical object in quantum mechanics is a complex vector space known as the \textit{Hilbert space}. It is customary to use the notation $\ket{\cdot }$, known as a \textit{ket}, for vectors in the Hilbert space. We also denote the conjugate transpose of $\ket{a}$ as $\bra{a}$, where $\bra{\cdot}$ is known as a \textit{bra}. The inner product of two kets $\ket{a}$ and $\ket{b}$ can be written as a \textit{bra-ket} $\braket{a}{b}.$

Operators on the Hilbert space can be thought of as complex-valued matrices. The most important operators are the Hamiltonian $H$, which governs the evolution of quantum systems, and the density matrix $\rho$, which describes the state of an open quantum system.

A particularly ubiquitous Hilbert space is that corresponding to particle number. The particle number Hilbert space has basis kets written $\ket{n}$, for $n\in \{0,1,2,\ldots\}$, where $\ket{n}$ represents a system with $n$ particles. This Hilbert space is ubiquitous; it can also be used to represent many 1d bound systems. In particular, this Hilbert space appears in most bosonic and continuous quantum systems.

In the particle number Hilbert space, there are special operators, the \textit{creation and annihilation operators}, which increase and decrease the number of particles, respectively. The creation operator $a^{\dagger}$ satisfies $a^{\dagger} \ket{n} = \sqrt{n+1} \ket{n+1}$ and the annihilation operator $a$ satisfies $a \ket{n} = \sqrt{n} \ket{n-1}$ with $a\ket{0}=0$.

The coherent state $\ket{\alpha}$ with a complex number $\alpha$ is defined as $\ket{\alpha} = e^{\alpha a^{\dagger} - \alpha^{*}a} \ket{0}$, where $e$ should be interpreted as the matrix exponential function.

Hilbert spaces of systems with multiple subsystems are tensor products of the subsystems' Hilbert spaces. Suppose we have two Hilbert spaces, $\mathcal{H}_1$ and $\mathcal{H}_2$. For every two kets $\ket{a}\in\mathcal{H}_1$ and $\ket{b}\in\mathcal{H}_2$, there exists a ket $\ket{a}\otimes\ket{b}\in \mathcal{H}_1\otimes \mathcal{H}_2$, where $\ket{a}\otimes\ket{b}$ is the tensor product of $\ket{a}$ and $\ket{b}$ and $\mathcal{H}_1\otimes \mathcal{H}_2$ is the tensor product space of Hilbert spaces $\mathcal{H}_1$ and $\mathcal{H}_2$. The inner product for tensor products of Hilbert spaces is defined as $\left(\bra{a}\otimes\bra{b}\right)\left(\ket{c}\otimes\ket{d}\right) = \braket{a}{c}\braket{b}{d}$, where $\bra{a}\otimes\bra{b}$ is the conjugate transpose of $\ket{a}\otimes\ket{b}$. If an operator $O_1$ acts on $\mathcal{H}_1$ and $O_2$ acts on $\mathcal{H}_2$, then $(O_1\otimes O_2)$ acts on $\mathcal{H}_1\otimes \mathcal{H}_2$ according to $(O_1\otimes O_2)\ket{a}\otimes \ket{b} = (O_1\ket{a})\otimes (O_2\ket{b}).$ Finally, note that we often use shorthands such as $O_1$ or $O_2$ to referer to $O_1\otimes 1$ or $1\otimes O_2$, respectively.

\subsection{Open Quantum System}

As discussed in Section \ref{sec: intro}, in an open quantum system, the state is described as a complex-valued, unit-trace positive definite matrix $\rho$, known as the \textit{density matrix}. The density matrix is a generalization of the wave function in the Schr\"odinger equation, which can be viewed as an ensemble of wave functions. 

A generic Markovian open quantum system has an evolution equation of the form 
\begin{equation}
    \label{eq:open}
    \dot\rho = \mathcal{L}\rho = -i\commutator{H}{\rho}+\mathcal{L}_{\text{loss}}\rho,
\end{equation}
where $H$ is the Hamiltonian matrix, $\mathcal{L}_{\text{loss}}$ is a dissipative operator, and $\commutator{\cdot}{\cdot}$ is the commutation operator between matrices, i.e., $[A, B] = AB - BA.$ Often, $H$ is composed of \textit{raising and lowering operators}, $a$ and $a^\dagger$. Here, $\mathcal{L}$ is a \textit{superoperator}; given a matrix $\rho$, it returns a new matrix $\mathcal{L}\rho.$ Eq.~\ref{eq:open} is a complex-valued high-dimensional differential equation, which is challenging to solve in general. 

Our work applies to open quantum systems with continuous degrees of freedom. Such systems include bosonic systems, which arise in a variety of contexts~\citep{cazalilla2011one,adesso2014continuous}. A bosonic particle, also known as a \emph{boson}, is a type of fundamental particle in quantum mechanics that has continuous degrees of freedom. Bosonic systems may be composed of multiple \textit{sites}, which are subsystems described by the particle number Hilbert space.

Simulating quantum systems with continuous variables introduces higher-dimensional complexity compared to those with discrete variables, such as spin systems. Even for a 1-site continuous variable system, there is infinite degree of freedom. In practice, one workaround is to truncate the infinite degree of freedom to some large finite degree $N$. Even with truncation, $k$ sites live in an exponentially-large-dimensional Hilbert space of size $N^k$, which is generally intractable to simulation. In contrast, our approach works with the infinite degree-of-freedom Hilbert space directly.

\subsection{Q Function Formulation}

The Husimi Q function~\citep{QuantumOpticsChapter1} provides an exact reformulation of Eq.~\ref{eq:open} into a probabilistic differential equation:
\begin{equation}
    \label{eq:openQ}
    \dot Q =  \tilde{\mathcal{L}} Q
\end{equation}
where $Q$ is the Husimi Q function, and $\tilde{\mathcal{L}}$ is the Q-function evolution operator including the effects of $H$ and $\mathcal{L}_{\text{loss}}$.

Mathematically, the Q function of $n$ sites is defined as\begin{equation*}
    Q(\vec{q}, \vec{p}) = Q(\vec{\alpha},\vec{\alpha}^*) = \frac{1}{\pi} \bra{\vec{\alpha}}\rho\ket{\vec{\alpha}},
\end{equation*} where $\vec{\alpha}=\vec{q}+i\vec{p}$ is a complex number, and \begin{equation*}
    \ket{\vec{\alpha}} = \ket{\alpha_1}\otimes\cdots\otimes\ket{\alpha_n}
\end{equation*} is a tensor product of coherent states.\footnote{Although $\alpha$ and $\alpha^*$ are both input to $Q$, it is customary in physics and complex analysis to write $Q(\alpha,\alpha^*)$ instead of $Q(\alpha).$} $Q(\vec{\alpha},\vec{\alpha}^*) \geq 0$ for any $\vec{\alpha}$ and $\int Q =1$, so $Q$ can be interpreted as a probability distribution in practice.

We use the notations $Q(\vec{q},\vec{p})$ and $Q(x)$ interchangeably.

To use the Q function formalism, we must convert between the $\rho$ and Q functions and obtain $\tilde{\mathcal{L}}$. We provide the key conversion formulas and the corresponding proofs in Appendix~\ref{app: Conversions}.

\subsection{Q-Flow representation: Flow-based Generative Models of Q function}

One important feature of our work is to represent the Q function with off-the-shelf flow-based generative models. This distinguishes our work from previous works~\citep{PhysRevLett.122.250503,PhysRevB.99.214306,PhysRevLett.122.250502,PhysRevLett.122.250501} that represent the high dimensional complex-valued density matrix using customized neural networks.
There are several advantages of our approach: i) we do not work with complex-valued functions, which could be complicated by the sign structure problem~\citep{westerhout2020generalization}. ii) Q-Flow is natural for systems with continuous degrees of freedom. iii) Q-Flow allows normalized probability modeling with exact sampling, which is important for solving high dimensional probabilistic PDEs with the stochastic Euler method.

\paragraph{Normalizing Flows.}
Normalizing flows are generative models for continuous probability distributions that provide both normalized probabilities and exact sampling---making them ideal for modeling the continuous Q function in our approach. Normalizing flows transform a simple initial density $p_X$ (often a unit-normal distribution) to a target density $p_Y$ (i.e.,\ the distribution that we want to model) via a sequence of invertible transformations~\citep{dinh2014nice,rezende2015variational}. The invertible transformations are usually parameterized by an invertible neural network architecture $y = f_\theta(x)$ with $x\sim p_X$ and $y\sim p_Y$. The target probability density is then given by
\begin{align*}
p_Y(y) &=p_X(f_\theta^{-1}(y))\left | \frac{\partial f_\theta^{-1}(y)}{\partial y} \right|.
\end{align*}
Many choices of $f_\theta$ are available, including affine coupling layers (RealNVP)~\citep{dinh2017density}, continuous normalizing flows (CNF) \citep{grathwohl2018scalable}, and convex potential flows (CP-Flow)~\citep{huang2021convex}. While RealNVP is the simplest to implement, affine coupling layers are less expressive than CNFs or CP-Flows, which are provably universal density estimators~\citep{huang2021convex}. Because of Equation \ref{eq:QtoRho}, we would like our flow to be infinitely differentiable, which is satisfied by the above flow architectures.

\begin{theorem}
    For a Q function from a given density matrix $\rho$, there exists a universal approximation with a Q-Flow representation.
\end{theorem}

\textit{Proof.} For any given density matrix $\rho$, there is a corresponding $Q_{\rho}$ which satisfies $Q_{\rho} \geq 0$ and $\int Q_{\rho} =1$.
Since it has been shown that normalization flow is a universal approximator of probability distribution~\citep{huang2021convex}, there exists a Q-Flow representation $Q_f$ such that it can be arbitrarily close to $Q_{\rho}$. \qed

\begin{theorem}
    For any local observable expected value to be computed with respect to $\rho$, there exists a Q-Flow representation which can compute the observable efficiently.
\end{theorem}

\textit{Proof.} We prove the single-site case here and the multi-site case follows from the tensor product structure of the Hilbert space. Consider the corresponding Q function $Q_{\rho}$ of $\rho$. Consider a local observable in the form of $O=a^ma^{\dagger n}+a^na^{\dagger m}$. WLOG, we can consider $O=a^ma^{\dagger n}$ and the other part can be done in a similar way.  its expectation$\langle O \rangle_{\rho} = \text{tr}(\rho a^m a^{\dagger n})$. Eq.~\ref{eq:obs} in Appendix shows that it can be equivalently computed by $\int (q+ip)^{m} (q-ip)^n Q_{\rho}(p,q)dp dq$, which is a polynomial moment of the Q function. Since normalization flow is a universal approximator, there exists a Q-Flow representation $Q_f$ can be arbitrarily close to $Q_{\rho}$, which implies that $\langle Q\rangle_f$ can be arbitrarily close to $\langle Q\rangle_{\rho}$. Even though computing $\langle Q\rangle_f = \sum_{(q,p)\sim Q_f} (q+ip)^{m} (q-ip)^n$ has stochastic fluctuation, the exact sampling nature of the flow-based model can suppress the statistical error, which will decay with increasing sample size $N_s$ as $\frac{1}{\sqrt{N_s}}$ due to the Central Limit Theorem. \qed

\subsection{Q-Flow Optimization: Stochastic Euler-KL Method}

In the previous section, we discuss the representation of the Q function with flow-based models. To solve the real-time dynamics given by Eq.~\ref{eq:openQ}, we further develop the high dimensional stochastic Euler-KL method.  

The algorithm represents the Q function at time $t$ with a flow-based model and iteratively updates the representation at the next time $t+dt$ based on the Euler method. It requires two copies of flow-based models for $Q^{t+dt}$ and $Q^{t}$. Based on the first-order Euler method with time step $dt$, Eq.~\ref{eq:openQ} yields \begin{equation}
    Q^{t+dt} = Q^{t} + \Tilde{\mathcal{L}}Q^{t}dt = (\mathrm{I}+\Tilde{\mathcal{L}}dt) Q^{t} \equiv Q_{\mathcal{L}}^{t}.
\end{equation}
Notice that $Q^{t+dt}$ represents the Q function that we obtain in the next time step.  At each learning step, we fix $Q^{t}$ and optimize the parameters $\theta$ in $Q^{t+dt}$ to match the above relation. Hence, we also denote $Q^{t+dt}$ by $Q_{\theta}^{t+dt}$. We train $Q_\theta^{t+dt}$ using the KL divergence loss function
\begin{equation}
    KL(Q_{\theta}^{t+dt} || Q_{\mathcal{L}}^{t}) = \int Q_{\theta}^{t+dt} \ln \frac{Q_{\theta}^{t+dt}}{Q_{\mathcal{L}}^{t}}.
    \label{eq:KL}
\end{equation}
The gradient of Eq.~\ref{eq:KL} can be derived with a control variance technique as follows (see Appendix for a derivation): 
\begin{equation}
    \frac{1}{N}\sum_{x \sim Q_{\theta}^{t+dt}} \left[\ln\frac{Q_{\theta}^{t+dt}(x)}{Q_{\mathcal{L}}^t(x)} - b\right]\nabla_\theta\ln Q_{\theta}^{t+dt}(x)
    \label{eq:KL_grad}
\end{equation}
where $b = \frac{1}{N} \sum_{x\sim Q^{t+dt}_{\theta}}\ln\frac{Q_{\theta}^{t+dt}(x)}{Q_{\mathcal{L}}^t(x)} $ is the baseline for control variance. 

The stochastic Euler-KL method is summarized in Algorithm.~\ref{alg:scheme}. We further provide an error bound by developing the analysis in  ~\citet{gutierrez2022real} to Q-Flow.

\begin{algorithm}[t]
\small
\caption{Stochastic Euler-KL Method}
\label{alg:scheme}
\begin{algorithmic}
\STATE\textbf{Input:} normalizing flow models for $Q_{\theta}^{t+dt}$ and $Q^t$, total time $T$, time step $dt$, $n_{iter}$, optimizer Adam.
 \STATE \textbf{Output:} Optimal parameters $\boldsymbol{\theta}^{*}$ at time step $t+dt$
  \STATE \textbf{Initialization:} Random $\boldsymbol{\theta} (t_0)$
  \FOR{$j$ in range($T/dt$)}
        \FOR{$i = 0$ to $n_{\mathrm{iter}}$}
        \STATE update $\theta$ using Eq.~\ref{eq:KL_grad} and optimizer Adam
        \ENDFOR
        \\
    $Q^t$ $\leftarrow$ $Q_{\theta^{*}}^{t+dt}$
    \ENDFOR 
\end{algorithmic}
\end{algorithm}

\begin{theorem}
    The global error $\es(t_n)$ of the n-step stochastic Euler method is bounded by $|\es_{E}(t_n)|+|\es_{NN}(t_n)|$, where $\es_{E}(t_n)$ is the global error of the exact Euler method and $\es_{NN}(t_n) = -P^{-1}\sum_{i=1}^{n} P^i r^{n+1-i}$ with $P = \mathrm{I}+\Tilde{\mathcal{L}}dt$ and $r^i$ being the $i$-th step stochastic Euler optimization error with neural network representation of the Q-Flow.
\end{theorem}

\textit{Proof.} $\es(t_n) = Q^{t_n} - Q_{NN}^{t_n} = (Q^{t_n} - Q_{E}^{t_n}) + (Q_{E}^{t_n} - Q_{NN}^{t_n}) \equiv \es_{E}(t_n) + \es_{NN}(t_n)$, where $Q_E^{t_n}$ and $Q_{NN}^{t_n}$ are the Q function from the exact Euler method and the neural network Q-Flow at time step $t_n$. By the triangular inequality, $|\es(t_n)|\leq |\es_{E}(t_n)|$ + $|\es_{NN}(t_n)|$. Since the Euler method is a first-order method, it has global error of order $O(dt)$ where $dt$ the time step. 

Denote the optimization error of Eq.~\ref{eq:KL} in time step $t_{n+1}$ as $r^{n+1}$, such that $Q_{NN}^{t_{n+1}} - PQ_{NN}^{t_n}=r^{n+1}$. It follows that $Q_E^{t_{n+1}}-\es_{NN}(t_{n+1}) - P(Q_E^{t_{n}}-\es_{NN}(t_n))=r^{n+1}$, which implies that $\es_{NN}(t_{n+1}) = P \es_{NN}(t_n)-r^{n+1}$ due to the cancellation of $Q_{E}^{t_{n+1}}-PQ_E^{t_n}$ from the exact Euler method. By induction, $\es_{NN}(t_n) = -P^{-1}\sum_{i=1}^{n} P^i r^{n+1-i}$. \qed

\textbf{Time Dependent Variational Principle (TDVP).} Instead of taking the gradient with respect to the KL divergence as Eq.~\ref{eq:KL_grad} shows,~\citet{reh2022variational} demonstrate that the minimization of Eq.~\ref{eq:KL} is equivalent to the time-dependent variational principle, which provides a nonlinear differential equation on the parameter space $\theta$ as follows.
\begin{equation}
    S_{kk'} \dot{\theta}_{k'} = F_k
    \label{eq:tdvp}
\end{equation}
where $S_{kk'} = \mathbb{E}[(\partial_{\theta_k}\ln Q)(\partial_{\theta_k'} \ln Q)]$ is the Fisher information matrix, and $F_k = \mathbb{E}[(\partial_{\theta_k}\ln Q)(\partial_t \ln Q)]$ with $\partial_t \ln Q = (\partial_t Q)/Q = (\tilde{\mathcal{L}}Q)/Q$.

\citet{reh2022variational} has only applied TDVP to solving classical PDEs. Under our Q-Flow approach, we can also apply TDVP to simulate open quantum dynamics. 

\textbf{Complexity Analysis.} Even though the stochastic Euler-KL method and the TDVP method are equivalent mathematically, they share different algorithmic complexity. TDVP requires solving the nonlinear differential equation in Eq.~\ref{eq:tdvp}, which requires explicitly inverting the Fisher information matrix $S_{kk'}$. Besides potential instability, this procedure has complexity scaling as $O(N^3)$ for explicit inversion, or $O(N^2)$ with the conjugate gradient approach, where $N$ is the number of parameters. This may limit its application for parameters beyond the orders of ten thousands. Meanwhile, the stochastic Euler method only requires first order optimization based on Eq.~\ref{eq:KL_grad}, the main cost of which comes from the number of optimization steps in each $dt$. 

\subsection{Q-Flow Initialization: Initial State Pretraining}\label{sec: pretraining}

Using a Q-Flow to simulate a quantum system requires initializing the flow to the correct starting Q function. For some simple initial states, we find that it is sufficient to simply make the initial state the prior for the flow and initialize the flow to the identity. However, we find that using more complex initial distributions as priors to a flow tends to hamper their ability to model a system's evolution. In these cases, we instead use the standard Gaussian prior, but we use a two-step process to pretrain the flow to match the initial distribution $Q_{\text{init}}$.

First, we sample from the desired initial distribution using the Metropolis-Hastings Monte Carlo method and update the flow parameters to minimize the negative log-likelihood $-\sum_{x\sim Q_{\text{init}}}\ln Q_\theta (x).$ This ensures that the model has some overlap with $Q_{\text{init}}$, which helps the next step's training algorithm converge more quickly.

Second, we sample from the flow and update the flow parameters to minimize the KL Loss, $KL(Q_{\text{init}}||Q_\theta)$. We compute the gradient according to \begin{equation}
    \nabla_\theta KL \approx -\frac{1}{N} \sum_{x\sim Q_{\theta}}\frac{Q_{\text{init}}(x)}{Q_{\theta}(x)}\nabla_\theta\ln Q_{\theta}(x).
\end{equation}

\begin{table}[t]
\centering

\scriptsize

\begin{tabular}{cccccc}

\toprule
\multicolumn{6}{c}{1-site}\\
\cmidrule(r){1-6}
& Q-Flow  & Q-Flow & & & \\
Time & Euler (ours) & TDVP (ours) & PINN & PS & FD \\
\cmidrule(r){1-1}
\cmidrule(r){2-2}
\cmidrule(r){3-3}
\cmidrule(r){4-4}
\cmidrule(r){5-5}
\cmidrule(r){6-6}

3 & 2.08e-3 & 5.11e-3 & 1.79e-1 & \textbf{3.47e-4} & 8.90e-4\\ 
6 & 5.10e-4 & 1.17e-3 & 1.84e-1 & \textbf{3.47e-4} & 9.01e-4\\ 
9 & \textbf{1.01e-4} & 2.16e-4 & 1.91e-1 & 3.47e-4 & 9.01e-4\\ 
12 & \textbf{1.68e-5} & 3.58e-5 & 1.91e-1 & 3.47e-4 & 9.01e-4\\ 
15 & 1.58e-5 & \textbf{5.55e-6} & 1.98e-1 & 3.47e-4 & 9.01e-4\\ 

\cmidrule(r){1-6}
\multicolumn{6}{c}{2-site}\\
\cmidrule(r){1-6}

3 & \textbf{3.91e-3} & 1.23e-2 & 1.00e0 & 1.83e-1 & 6.12e-2\\ 
6 & \textbf{1.91e-3} & 4.66e-3 & 1.00e0 & 1.82e-1 & 6.09e-2\\ 
9 & \textbf{7.59e-4} & 1.77e-3 & 1.00e0 &  1.81e-1 & 6.09e-2\\ 
12 & \textbf{2.92e-4} & 6.21e-4 & 1.00e0 &  1.81e-1 & 6.09e-2\\ 
15 & \textbf{1.47e-4} & 2.05e-4 & 1.00e0 &  1.81e-1 & 6.09e-2\\ 

\cmidrule(r){1-6}
\multicolumn{6}{c}{20-site}\\
\cmidrule(r){1-6}

3 & \textbf{9.94e-2} & 1.08e-1 & 2.17e31 &  - & -\\ 
6 & \textbf{3.29e-2} & 4.10e-2 & 2.38e30 &  - & -\\ 
9 & \textbf{2.02e-2} & 2.44e-2 & 1.34e29 &  - & -\\ 
12 & \textbf{1.46e-2} & 1.68e-2 & 1.46e28 &  - & -\\ 
15 & \textbf{1.07e-2} & 1.23e-2 & 7.07e26 &  - & -\\ 

\bottomrule
\end{tabular}

\caption{$L_1[Q_{\text{sim}},Q_{\text{exact}}]$ for each simulation method over time. For each row, we mark the best result in bold.}
\label{table: Fokker-Planck L1 Loss}
\end{table}

\section{Experiments}

For our experiments, we focus on two types of open quantum systems: dissipative harmonic oscillators and dissipative bosonic systems. We test on dissipative harmonic oscillators because they have an analytic solution, which makes them useful for benchmarking high-dimensional PDE solvers beyond the limits of conventional solvers. We then test on dissipative bosonic systems because they are commonly studied and of practical use in physics. 

In these experiments, we compare Euler and TDVP methods to PINNs, Pseudo-spectral solvers, Finite Difference solvers, and stochastic solvers. Although we do not develop the TDVP method, we propose a method to apply it to open bosonic quantum systems. As such, we sometimes describe the Euler and TDVP methods as ``our methods."

For our experiments, we use Affine Coupling Flows and Convex-Potential Flows for the Euler and TDVP methods. Affine Coupling Flows are fast but less expressive, so we use them for the dissipative harmonic oscillator experiments. Convex Potential Flows are slow but more expressive, so we use them for problems involving more complex Q functions.

To run our experiments~\cite{code}, we use the Jax library \cite{jax2018github} for Euler and TDVP methods. We make use of the \hyperlink{https://github.com/ChrisWaites/jax-flows}{jax-flows} library. To implement the TDVP method, we make use of the NetKet library \cite{netket2:2019,netket3:2022} and its Stochastic Reconfiguration~\citep{sorella1998green,sorella2001generalized} feature, which is mathematically equivalent to TDVP. For distributed training, NetKet uses the mpi4jax package \cite{mpi4jax:2021}. For PINNs, we use the \hyperlink{https://github.com/mathLab/PINA}{PINA} library, which is built on top of PyTorch. Finally, for the other three baselines we use Julia~\citep{rackauckas2017differentialequations}.

Further explanation of the observables chosen and their significance can be found in Appendix \ref{app: why obs}. More details about the Normalizing Flow models we use are provided in Appendix \ref{app:Flow details}. More details about experimental setup, hyperparameters, and baselines are provided in Appendix \ref{app: Experiment details}.

\subsection{Dissipative Harmonic Oscillator}\label{sec: DHO}

\textbf{Experimental Setup.} The multi-site dissipative harmonic oscillator evolves according to Equation \ref{eq:open} with Hamiltonian \cite{QuantumOpticsChapter1} $H = \sum_j \omega_j a_j^\dagger a_j$ and loss term \begin{align}
    \begin{split}
        \mathcal{L}_{\text{loss}}\rho = &\sum_j \gamma_j\left[\frac{1}{2}(2a_j\rho a_j^\dagger -a_j^\dagger a_j\rho -\rho a_j^\dagger a_j)\right.\\
        &+ \bar{n}_j(a_j\rho a_j^\dagger + a_j^\dagger\rho a_j -a_j^\dagger a_j\rho -\rho a_ja_j^\dagger)\bigg].
    \end{split}
\end{align} Here, $j$ labels what we will call \textit{sites}. Converting to the Q function formalism gives \cite{QuantumOpticsChapter4} \begin{align*}
    \tilde{\mathcal{L}} = &\sum_j \left[\gamma_j + \frac{1}{4}\gamma_j(\bar{n}_j+1)\left(\partialderivative{{}^2}{q_j^2} + \partialderivative{{}^2}{p_j^2}\right)\right.\\
    &\;\;+\left(\frac{\gamma_j}{2}q_j-\omega_jp_j\right)\partialderivative{}{q_j}+\left(\frac{\gamma_j}{2}p_j+\omega_jq_j\right)\partialderivative{}{p_j}\Bigg].
\end{align*} We test the simulation methods on three problems of increasing dimensionality: a 1-site system, a 2-site system, and a 20-site system. For each system, we use a coherent state initial condition, which corresponds to a Gaussian with variance $1/2.$ We center the Gaussian at $(-1,\ldots,-1)$. As time passes this Gaussian spirals toward the origin and changes its standard deviation. To make the simulation more challenging, for every site $j$ we uniformly sample the system's parameters $\bar{n}_j\in [3,7),$ $\gamma_j\in [0.5,1.5),$ and $\omega_j\in [0.5,1.5).$ See Appendix \ref{app: Fokker-Planck Parameters} for more details about the choice of system parameters.

\begin{table}[t]
\centering

\label{table: Fokker-Planck Rho L2 Loss}

\small
\begin{tabular}{cccc}

\toprule
Time & Q-Flow (Euler)  & Q-Flow (TDVP) & PINN\\
\cmidrule(r){1-1}
\cmidrule(r){2-2}
\cmidrule(r){3-3}
\cmidrule(r){4-4}

3 & $\mathbf{1.30\cdot 10^{-7}}$ & $8.18\cdot 10^{-7}$ & $2.43\cdot 10^{-3}$ \\ 
6 & $\mathbf{7.64\cdot 10^{-9}}$ & $4.06\cdot 10^{-8}$ & $2.66\cdot 10^{-3}$\\ 
9 & $\mathbf{3.16\cdot 10^{-10}}$ & $1.38\cdot 10^{-9}$ & $2.68\cdot 10^{-3}$\\ 
12 & $\mathbf{7.49\cdot 10^{-12}}$ & $3.79\cdot 10^{-11}$ & $2.34\cdot 10^{-3}$\\ 
15 & $4.30\cdot 10^{-12}$ & $\mathbf{9.21\cdot 10^{-13}}$ & $1.89\cdot 10^{-3}$\\

\bottomrule
\end{tabular}

\caption{$L_2$ loss for each simulation method's density matrix over time for the 1-site system. We mark each row's best result in bold.}
\end{table}

\textbf{Metrics.} To evaluate performance, we compute the $L_1$ Loss between each simulation and the exact distribution:\begin{equation}
    \label{eq: L1}
    \begin{aligned}
        L_1[Q_{\text{sim}},Q_{\text{exact}}] &\equiv \int \text{d}^d x \; |Q_{\text{sim}}(x)-Q_{\text{exact}}(x)|\\
        &\approx \frac{1}{N}\sum_{x {\sim}Q_{\text{exact}}} \left|\frac{Q_{\text{sim}}(x)}{Q_{\text{exact}}(x)} - 1\right|.
    \end{aligned}
\end{equation} Although the $L_1$ Loss is a useful metric, it is also illustrative to examine observables of the system. One observable is the centroid, $\mathbb{E}[\vec{x}] \approx \frac{1}{N}\sum_{x {\sim}Q_{\text{sim}}} \vec{x}.$ With more sites, we cannot easily plot the centroid trajectory, so instead we compute the centroid's distance from the origin, $\norm{\mathbb{E}[\vec{x}]}$.

\begin{figure}
    \centering
    
    \includegraphics[width=\linewidth]{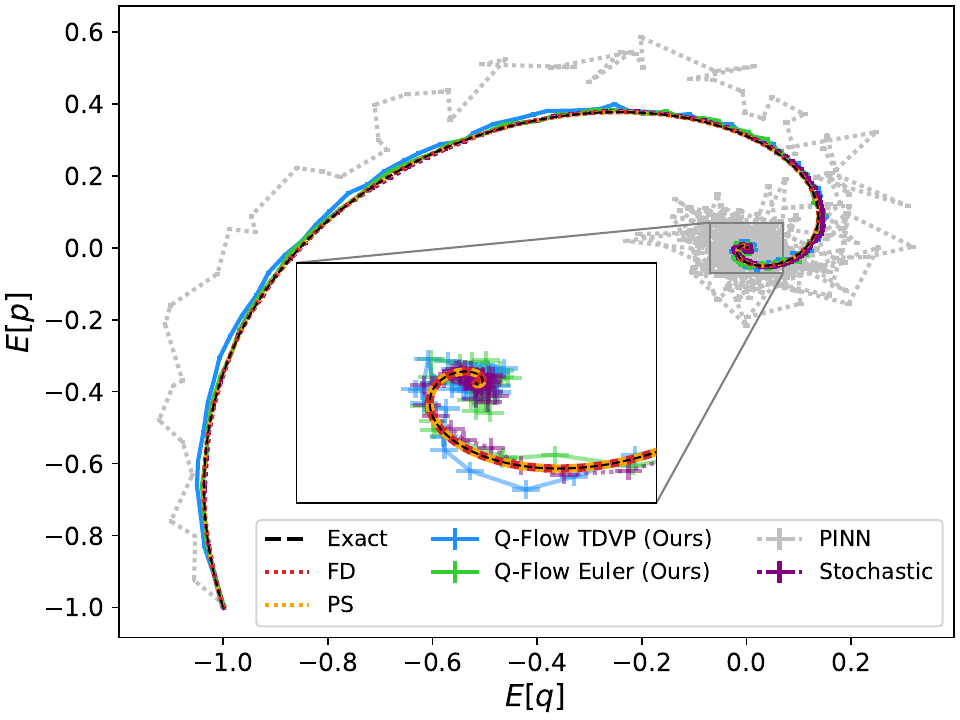}
    
    \caption{The trajectory of the centroids of the simulated distributions. The PINN baseline is excluded from the inset. Error bars are included for all but the FD and PS methods but are small.}
    
    \label{fig: Fokker-Planck Centroid}
\end{figure}

\begin{figure*}
    \centering
    
    \includegraphics[width=\linewidth]{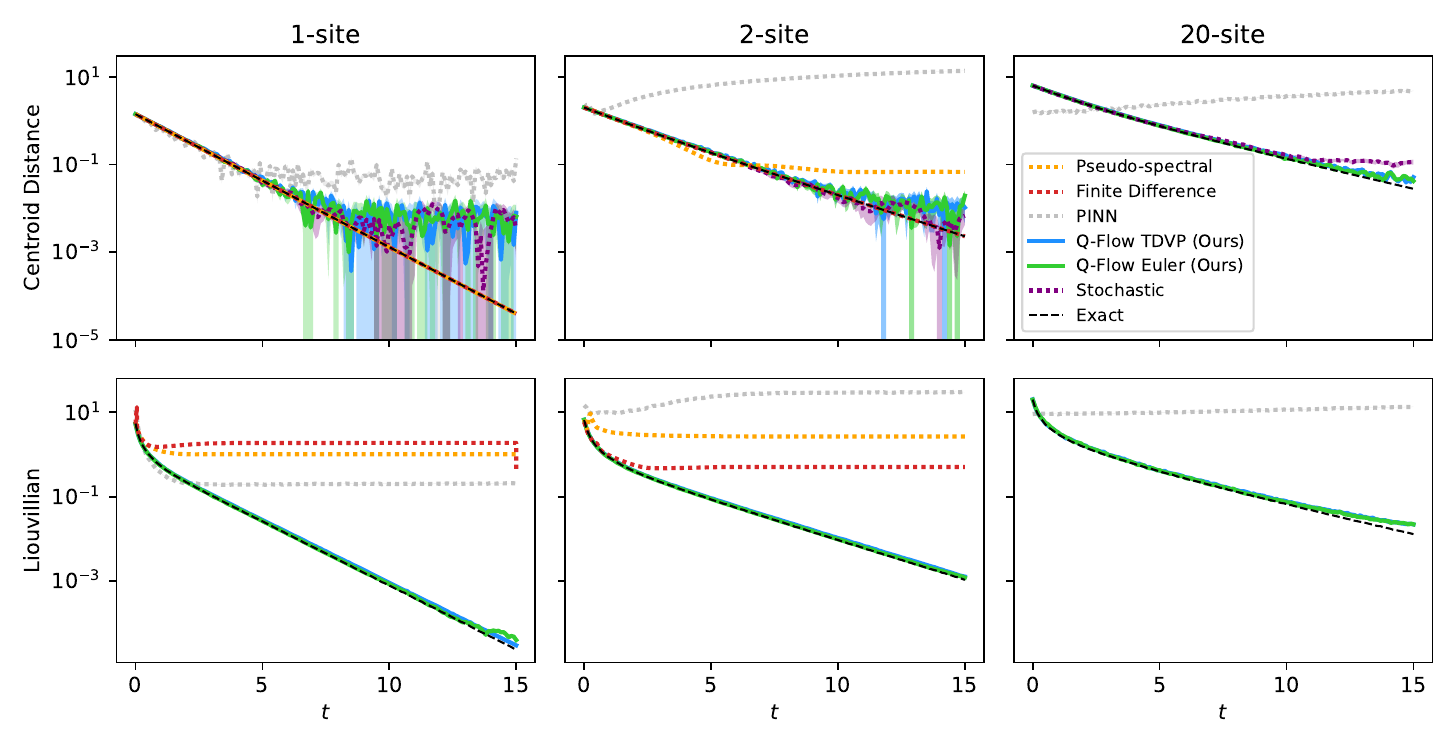}
    
    \caption{The simulated evolution of two observables for 1-site, 2-site, and 20-site dissipative harmonic oscillators. Error bars are included for all but the Finite Difference and Pseudo-spectral results but are small for most observables.}
    
    \label{fig: Fokker-Planck Observables}
\end{figure*}

Additionally, we compute the \textit{Liouvillian loss}, \begin{equation*}
    \int dx\; \left|[\tilde{\mathcal{L}}Q](x)\right| = \mathbb{E}[|\tilde{\mathcal{L}}Q|/Q].
\end{equation*} The Liouvillian loss measures the magnitude of the dynamics relative to the distribution, an indicator of how perturbed the system is from equilibrium.

For the Euler and TDVP methods, we sample directly from the flow to compute expected values. For PINNs, we use Markov chain Monte Carlo (MCMC) to obtain samples. For pseudo-spectral results, we compute expected values by summing over the grid and scaling by $Q$. 

Finally, for the 1-site system, we compute the first 4x4 block of the density matrix according to Equation \ref{eq:QtoRho} and compute its $L_2$ distance from the exact density matrix:\begin{equation}\label{eq: L2}
    L_2[\rho_{\text{pred}},\rho_{\text{exact}}] = \sum_{1\le i,j\le 4}\left|(\rho_{\text{pred}})_{ij}-(\rho_{\text{exact}})_{ij}\right|^2.
\end{equation} Because the density matrix is the standard parametrization of a quantum system, this comparison is another useful benchmark for performance. Equation \ref{eq:QtoRho} requires spatial derivatives of the Q-function, so we only compute this loss for solvers that return spatially differentiable Q functions.

\textbf{Results and Discussion.} 
Table \ref{table: Fokker-Planck L1 Loss} shows the $L_1$ Loss between each simulation and the exact distribution for a number of simulation times. Although we do not include error bounds in the table for ease of viewing, the error is usually at least an order of magnitude smaller than the $L_1$ Loss (see Appendix.~\ref{app:additional experiment}). Error bounds for the pseudo-spectral and finite-difference results (standard solvers) are not computed because these methods are deterministic. We exclude standard solvers from the 20-site system because a grid size of only 10 would require storing at least $10^{40}$ values.

Both the Euler and TDVP methods have extremely low $L_1$ Loss. Both methods perform better than the standard solvers in the 2-site case and in the later times of the 1-site case. Increasing the number of sites, we find that the Euler and TDVP methods continue to perform well while PINNs and standard solvers struggle. Standard solvers cannot simulate the 20-site system due to the curse of dimensionality, and while PINNs can in principle simulate the system, in practice they perform extremely poorly. On the other hand, both the Euler and TDVP methods still consistently report low fidelities. Finally, note that the Euler method has a consistently lower $L_1$ loss than the TDVP method.

Figure \ref{fig: Fokker-Planck Centroid} shows the trajectory of each simulation method's centroid for the 1-site case. Once again, the Euler and TDVP methods both closely match the exact evolution, and the Euler method performs slightly better in general. On the other hand, the PINN solution exhibits consistently biased and rapidly fluctuating estimates of the centroid. As expected, the standard solvers closely track the exact trajectory. However, we note that although our methods appear to match the exact results less accurately, the large error bars in the cutout demonstrate that this is in large part due to sampling error. We could have computed the centroid for the Euler and TDVP methods using grid integration as with the pseudo-spectral method, but we instead choose to use sampling because this better generalizes to higher dimensions. The stochastic method performs comparably to our Euler and TDVP methods, but we note that it applies to a restricted subset of diffusion-type PDEs. Additionally, the stochastic method cannot provide exact values of the Q function, which makes it challenging to evaluate other observables such as the Liouvillian loss or $L_1$ and $L_2$ losses.

Figure \ref{fig: Fokker-Planck Observables} shows the evolution of the centroid distance and the Liouvillian Loss for all three problems. Euler and TDVP closely match the exact evolution of the two observables. Although the two methods' estimates of the centroid distance begin to diverge from the exact centroid distance at around time 10, once again the large error bars demonstrate that this is due to error in the sampling estimate. Although the PINN centroid distance also begins to diverge from the desired value, the small error bars for this estimate suggest that the deviation does not come from sampling error.

The Euler and TDVP methods' Liouvillian losses  decrease consistently. At around time 15, the Euler Liouvillian loss jumps slightly. This jump occurs at a Liouvillian loss below $10^{-7}$, so the simulation is still likely precise enough for most applications. The Euler method's performance can likely be improved by increasing the number of fitting steps per time step and by decreasing the step size. In practice, we find that decreasing the step size improves both the Euler and TDVP methods' performance. Interestingly, the standard solvers provide very poor estimates of the Liouvillian loss. We suspect that this is due to error in numerical derivatives.

Finally, note that unlike the other methods, our methods continue to correctly simulate the system for large numbers of sites. It is only toward the end of time evolution in the 20-site case that our methods begin to show some deviation from the exact observables. Again, this can likely be reduced by decreasing the step size and taking more samples. Interestingly, the stochastic method appears to diverge slightly more than our methods in the 20-site case.

Table \ref{table: Fokker-Planck Rho L2 Loss} shows the $L_2$ loss from Equation \ref{eq: L2}. The Euler and TDVP methods have extremely low losses, with the Euler method performing slightly better. The PINN $L_2$ loss, while low, is much larger than the Euler and TDVP methods.

\begin{figure}[t]
    \centering

    \includegraphics[width=
\linewidth]{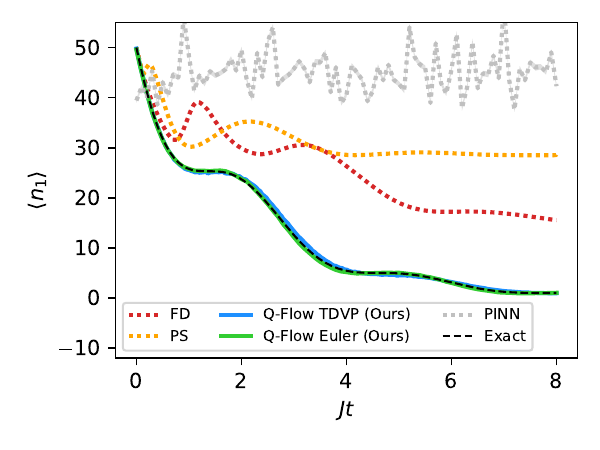}
    
    \caption{2-site dissipative bosonic experiment: simulated $\langle n_1\rangle$.}
    
    \label{fig: Bose-Hubbard 1}
\end{figure}

\subsection{Dissipative Bosonic Model} \label{sec: BH}

\textbf{Experimental Setup.} The dissipative bosonic model is a frequently studied open quantum system \cite{DecayFragmendationBoseHubbard,BERG2019239}. We test our methods on this model because it has a more complex evolution equation with rich real-world applicabilitions.

The dissipative bosonic model we use has \cite{DecayFragmendationBoseHubbard} $H = -J\sum_j\left(a_{j+1}^\dagger a_j+a_j^\dagger a_{j+1}\right)$ and \begin{align}
    \mathcal{L}_{\text{loss}}\rho = -&\frac{1}{2}\sum_j \gamma_j\left(n_j\rho + \rho n_j - 2a_j\rho a_j^\dagger\right)
\end{align} where $n_j = a_j^\dagger a_j$ and $j$ enumerates the sites. Converting to the Q function formalism gives that $\tilde{\mathcal{L}}$ is
\begin{equation*}
    \begin{aligned}
        &\sum_j \gamma_j\left(\frac{1}{4}\left(\partialderivative{{}^2}{q_j^2}+\partialderivative{{}^2}{p_j^2}\right)+\frac{1}{2}\left(q_j\partialderivative{}{q_j}+p_j\partialderivative{}{p_j}+1\right)\right)\\
        &+J\sum_j \Bigg(p_{j+1}\partialderivative{}{q_j}-q_{j+1}\partialderivative{}{p_j}
        +p_{j}\partialderivative{}{q_{j+1}}-q_{j}\partialderivative{}{p_{j+1}}\Bigg).
    \end{aligned}
\end{equation*} Following Figure 3 of \cite{BoseHubbardMethodsExamples}, we consider a 2-site system with $J=1,$ $U=0,$ and $\gamma = [1,0]$.  We simulate the evolution of an antisymmetric Bose-Einstein Condensate (BEC) with 50 particles in each site, which has a Q function $Q(q_1,p_1,q_2,p_2)$ given by \begin{equation*}
    Q = \frac{\left[(q_1-q_2)^2+(p_1-p_2)^2\right]^{100}}{\pi^2\cdot 2^{100}\cdot 100!} e^{-(q_1^2+p_1^2+q_2^2+p_2^2)}.
\end{equation*} Because of the complex multimodal initial distribution, we use the Convex Potential Flow for these experiments. We pretrain the flow as described in Section \ref{sec: pretraining}. 

\textbf{Metric.} For this system, we compute the observable $\langle n_1\rangle \approx \frac{1}{N}\sum_{(\vec{q},\vec{p}){\sim}Q_{\text{sim}}} (q_1^2+p_1^2-1)$ because it's exact evolution is given in \cite{BoseHubbardMethodsExamples}.

\textbf{Results and Discussion.} We show the simulated evolution of $\langle n_1\rangle$ in Figure \ref{fig: Bose-Hubbard 1}. The existence of the $J$ term is responsible for the oscillations shown because it causes the two sites to exchange particles, which could cause challenges for simulations. Even so, both the Euler and TDVP methods closely match the exact evolution, demonstrating the wide applicability of our methods. 

\section{Conclusion}
In this work, we made an important contribution to the problem of simulating open quantum systems. We used a reformulation of the density matrix to the Husimi Q function, which allowed us to study open quantum systems as an evolution of a probability distribution under dynamics, described by a partial differential equation that we derive for each system. This allowed us to establish a direct connection between simulating continuous or bosonic open quantum systems and the rich literature on generative models in standard machine learning. With off-the-shelf normalizing flows, Affine Coupling Flows and Convex Potential Flows, and a new efficient method for solving high-dimensional PDEs, Euler-KL, we established Q-Flow, a new and efficient approach to simulation of open quantum systems.

We compared Q-Flow to the state-of-the-art numerical and deep learning approaches on two important systems to the field, the dissipative harmonic oscillator and dissipative bosonic models. We established superior performance across the board, especially for large system dimensionality.

We believe the significance of our results is twofold. On one hand, Q-Flow's accurate simulation of open quantum systems can be further developed to aid progress in fundamental physics and engineering applications, such as superconductors and quantum computers. On the other hand, through our reformulation from evolving the density matrix to evolving the Q function, we shifted the modeling challenges from the curse of dimensionality to the accurate evolution of a high-dimensional deep generative model. Q-Flow can aid progress in evolving probability distributions under PDE dynamics and inspire future work on deep generative models.

\section{Acknowledgements}
The authors acknowledge helpful discussions with Zhuo Chen and Matija Medvidović. The authors acknowledge support from the National Science Foundation under Cooperative Agreement PHY-2019786 (The NSF AI Institute for Artificial Intelligence and Fundamental Interactions, \url{http://iaifi.org/}). This material is based upon work supported by the U.S. Department of Energy, Office of Science, National Quantum Information Science Research Centers, Co-design Center for Quantum Advantage (C2QA) under contract number DE-SC0012704. This work is also work supported in part by
the Air Force Office of Scientific Research under the award number FA9550-21-1-0317. P.Y.~Lu is grateful for the support of the Eric and Wendy Schmidt AI in Science Postdoctoral Fellowship at the University of Chicago.
\bibliography{biblio}

\begin{thebibliography}{57}
\providecommand{\natexlab}[1]{#1}
\providecommand{\url}[1]{\texttt{#1}}
\expandafter\ifx\csname urlstyle\endcsname\relax
  \providecommand{\doi}[1]{doi: #1}\else
  \providecommand{\doi}{doi: \begingroup \urlstyle{rm}\Url}\fi

\bibitem[Adesso et~al.(2014)Adesso, Ragy, and Lee]{adesso2014continuous}
Adesso, G., Ragy, S., and Lee, A.~R.
\newblock Continuous variable quantum information: Gaussian states and beyond.
\newblock \emph{Open Systems \& Information Dynamics}, 21\penalty0
  (01n02):\penalty0 1440001, 2014.

\bibitem[Amari(1996)]{amari1996neural}
Amari, S.-i.
\newblock Neural learning in structured parameter spaces-natural riemannian
  gradient.
\newblock \emph{Advances in neural information processing systems}, 9, 1996.

\bibitem[Amari(1998)]{amari1998natural}
Amari, S.-I.
\newblock Natural gradient works efficiently in learning.
\newblock \emph{Neural computation}, 10\penalty0 (2):\penalty0 251--276, 1998.

\bibitem[Barreiro et~al.(2011)Barreiro, M{\"u}ller, Schindler, Nigg, Monz,
  Chwalla, Hennrich, Roos, Zoller, and Blatt]{barreiro2011open}
Barreiro, J.~T., M{\"u}ller, M., Schindler, P., Nigg, D., Monz, T., Chwalla,
  M., Hennrich, M., Roos, C.~F., Zoller, P., and Blatt, R.
\newblock An open-system quantum simulator with trapped ions.
\newblock \emph{Nature}, 470\penalty0 (7335):\penalty0 486--491, 2011.

\bibitem[Berg \& Nyström(2019)Berg and Nyström]{BERG2019239}
Berg, J. and Nyström, K.
\newblock Data-driven discovery of {PDEs} in complex datasets.
\newblock \emph{Journal of Computational Physics}, 384:\penalty0 239--252,
  2019.
\newblock ISSN 0021-9991.
\newblock \doi{https://doi.org/10.1016/j.jcp.2019.01.036}.
\newblock URL
  \url{http://www.sciencedirect.com/science/article/pii/S0021999119300944}.

\bibitem[Bradbury et~al.(2018)Bradbury, Frostig, Hawkins, Johnson, Leary,
  Maclaurin, Necula, Paszke, Vander{P}las, Wanderman-{M}ilne, and
  Zhang]{jax2018github}
Bradbury, J., Frostig, R., Hawkins, P., Johnson, M.~J., Leary, C., Maclaurin,
  D., Necula, G., Paszke, A., Vander{P}las, J., Wanderman-{M}ilne, S., and
  Zhang, Q.
\newblock {JAX}: composable transformations of {P}ython+{N}um{P}y programs,
  2018.
\newblock URL \url{http://github.com/google/jax}.

\bibitem[Carleo \& Troyer(2017)Carleo and Troyer]{doi:10.1126/science.aag2302}
Carleo, G. and Troyer, M.
\newblock Solving the quantum many-body problem with artificial neural
  networks.
\newblock \emph{Science}, 355\penalty0 (6325):\penalty0 602--606, 2017.
\newblock \doi{10.1126/science.aag2302}.

\bibitem[Carleo et~al.(2019)Carleo, Choo, Hofmann, Smith, Westerhout, Alet,
  Davis, Efthymiou, Glasser, Lin, Mauri, Mazzola, Mendl, van Nieuwenburg,
  O'Reilly, Th{\'e}veniaut, Torlai, Vicentini, and Wietek]{netket2:2019}
Carleo, G., Choo, K., Hofmann, D., Smith, J. E.~T., Westerhout, T., Alet, F.,
  Davis, E.~J., Efthymiou, S., Glasser, I., Lin, S.-H., Mauri, M., Mazzola, G.,
  Mendl, C.~B., van Nieuwenburg, E., O'Reilly, O., Th{\'e}veniaut, H., Torlai,
  G., Vicentini, F., and Wietek, A.
\newblock Netket: A machine learning toolkit for many-body quantum systems.
\newblock \emph{SoftwareX}, pp.\  100311, 2019.
\newblock \doi{10.1016/j.softx.2019.100311}.
\newblock URL
  \url{http://www.sciencedirect.com/science/article/pii/S2352711019300974}.

\bibitem[Carmichael(1999{\natexlab{a}})]{QuantumOpticsChapter1}
Carmichael, H.~J.
\newblock \emph{Dissipation in Quantum Mechanics: The Master Equation
  Approach}, pp.\  1--28.
\newblock Springer Berlin Heidelberg, Berlin, Heidelberg, 1999{\natexlab{a}}.
\newblock ISBN 978-3-662-03875-8.
\newblock \doi{10.1007/978-3-662-03875-8_1}.
\newblock URL \url{https://doi.org/10.1007/978-3-662-03875-8_1}.

\bibitem[Carmichael(1999{\natexlab{b}})]{QuantumOpticsChapter4}
Carmichael, H.~J.
\newblock \emph{Quantum---Classical Correspondence for the Electromagnetic
  Field II: P, Q, and Wigner Representations}, pp.\  101--145.
\newblock Springer Berlin Heidelberg, Berlin, Heidelberg, 1999{\natexlab{b}}.
\newblock ISBN 978-3-662-03875-8.
\newblock \doi{10.1007/978-3-662-03875-8_4}.
\newblock URL \url{https://doi.org/10.1007/978-3-662-03875-8_4}.

\bibitem[Cazalilla et~al.(2011)Cazalilla, Citro, Giamarchi, Orignac, and
  Rigol]{cazalilla2011one}
Cazalilla, M.~A., Citro, R., Giamarchi, T., Orignac, E., and Rigol, M.
\newblock One dimensional bosons: From condensed matter systems to ultracold
  gases.
\newblock \emph{Reviews of Modern Physics}, 83\penalty0 (4):\penalty0 1405,
  2011.

\bibitem[Chen et~al.(2022)Chen, Luo, Hu, and Clark]{chen2022simulating}
Chen, Z., Luo, D., Hu, K., and Clark, B.~K.
\newblock Simulating 2+ 1d lattice quantum electrodynamics at finite density
  with neural flow wavefunctions.
\newblock \emph{arXiv preprint arXiv:2212.06835}, 2022.

\bibitem[Courant et~al.(1928)Courant, Friedrichs, and Lewy]{courant1928uber}
Courant, R., Friedrichs, K., and Lewy, H.
\newblock Über die partiellen differenzengleichungen der mathematischen
  physik.
\newblock \emph{Mathematische Annalen}, 100:\penalty0 32--74, 1928.
\newblock \doi{10.1007/BF01448839}.
\newblock URL \url{https://doi.org/10.1007/BF01448839}.

\bibitem[Dinh et~al.(2014)Dinh, Krueger, and Bengio]{dinh2014nice}
Dinh, L., Krueger, D., and Bengio, Y.
\newblock Nice: Non-linear independent components estimation.
\newblock \emph{arXiv preprint arXiv:1410.8516}, 2014.

\bibitem[Dinh et~al.(2017)Dinh, Sohl-Dickstein, and Bengio]{dinh2017density}
Dinh, L., Sohl-Dickstein, J., and Bengio, S.
\newblock Density estimation using real {NVP}.
\newblock In \emph{International Conference on Learning Representations}, 2017.
\newblock URL \url{https://openreview.net/forum?id=HkpbnH9lx}.

\bibitem[Dugan et~al.(2023)Dugan, Lu, Dangovski, Luo, and
  Solja{\v{c}}i{\'c}]{code}
Dugan, O., Lu, P.~Y., Dangovski, R., Luo, D., and Solja{\v{c}}i{\'c}, M.
\newblock The code repository is going to be made public in arxiv and the
  related materials are available on reasonable request from the corresponding
  author.
\newblock \emph{arXiv preprint arXiv:2302.12235}, 2023.

\bibitem[Fornberg(1998)]{fornberg1998practical}
Fornberg, B.
\newblock \emph{A practical guide to pseudospectral methods}.
\newblock Number~1. Cambridge university press, 1998.

\bibitem[Grathwohl et~al.(2019)Grathwohl, Chen, Bettencourt, and
  Duvenaud]{grathwohl2018scalable}
Grathwohl, W., Chen, R. T.~Q., Bettencourt, J., and Duvenaud, D.
\newblock Scalable reversible generative models with free-form continuous
  dynamics.
\newblock In \emph{International Conference on Learning Representations}, 2019.
\newblock URL \url{https://openreview.net/forum?id=rJxgknCcK7}.

\bibitem[Guti{\'e}rrez \& Mendl(2022)Guti{\'e}rrez and
  Mendl]{gutierrez2022real}
Guti{\'e}rrez, I.~L. and Mendl, C.~B.
\newblock Real time evolution with neural-network quantum states.
\newblock \emph{Quantum}, 6:\penalty0 627, 2022.

\bibitem[Hartmann \& Carleo(2019)Hartmann and Carleo]{PhysRevLett.122.250502}
Hartmann, M.~J. and Carleo, G.
\newblock Neural-network approach to dissipative quantum many-body dynamics.
\newblock \emph{Phys. Rev. Lett.}, 122:\penalty0 250502, Jun 2019.
\newblock \doi{10.1103/PhysRevLett.122.250502}.
\newblock URL \url{https://link.aps.org/doi/10.1103/PhysRevLett.122.250502}.

\bibitem[{Hendrycks} \& {Gimpel}(2016){Hendrycks} and {Gimpel}]{GELU}
{Hendrycks}, D. and {Gimpel}, K.
\newblock {Gaussian Error Linear Units (GELUs)}.
\newblock \emph{arXiv e-prints}, art. arXiv:1606.08415, June 2016.
\newblock \doi{10.48550/arXiv.1606.08415}.

\bibitem[Hermann et~al.(2020)Hermann, Sch{\"a}tzle, and No{\'e}]{Hermann2020}
Hermann, J., Sch{\"a}tzle, Z., and No{\'e}, F.
\newblock Deep-neural-network solution of the electronic schr{\"o}dinger
  equation.
\newblock \emph{Nature Chemistry}, 12\penalty0 (10):\penalty0 891--897, Oct
  2020.
\newblock ISSN 1755-4349.
\newblock \doi{10.1038/s41557-020-0544-y}.

\bibitem[Huang et~al.(2021)Huang, Chen, Tsirigotis, and
  Courville]{huang2021convex}
Huang, C.-W., Chen, R. T.~Q., Tsirigotis, C., and Courville, A.
\newblock Convex potential flows: Universal probability distributions with
  optimal transport and convex optimization.
\newblock In \emph{International Conference on Learning Representations}, 2021.
\newblock URL \url{https://openreview.net/forum?id=te7PVH1sPxJ}.

\bibitem[Häfner \& Vicentini(2021)Häfner and Vicentini]{mpi4jax:2021}
Häfner, D. and Vicentini, F.
\newblock mpi4jax: Zero-copy mpi communication of jax arrays.
\newblock \emph{Journal of Open Source Software}, 6\penalty0 (65):\penalty0
  3419, 2021.
\newblock \doi{10.21105/joss.03419}.
\newblock URL \url{https://doi.org/10.21105/joss.03419}.

\bibitem[{Kingma} \& {Dhariwal}(2018){Kingma} and {Dhariwal}]{GLOW}
{Kingma}, D.~P. and {Dhariwal}, P.
\newblock {Glow: Generative Flow with Invertible 1x1 Convolutions}.
\newblock \emph{arXiv e-prints}, art. arXiv:1807.03039, July 2018.
\newblock \doi{10.48550/arXiv.1807.03039}.

\bibitem[Kordas et~al.(2013)Kordas, Wimberger, and
  Witthaut]{DecayFragmendationBoseHubbard}
Kordas, G., Wimberger, S., and Witthaut, D.
\newblock Decay and fragmentation in an open bose-hubbard chain.
\newblock \emph{Phys. Rev. A}, 87:\penalty0 043618, Apr 2013.
\newblock \doi{10.1103/PhysRevA.87.043618}.
\newblock URL \url{https://link.aps.org/doi/10.1103/PhysRevA.87.043618}.

\bibitem[{Kordas} et~al.(2015){Kordas}, {Witthaut}, {Buonsante}, {Vezzani},
  {Burioni}, {Karanikas}, and {Wimberger}]{BoseHubbardMethodsExamples}
{Kordas}, G., {Witthaut}, D., {Buonsante}, P., {Vezzani}, A., {Burioni}, R.,
  {Karanikas}, A.~I., and {Wimberger}, S.
\newblock {The dissipative Bose-Hubbard model}.
\newblock \emph{European Physical Journal Special Topics}, 224\penalty0
  (11):\penalty0 2127--2171, November 2015.
\newblock \doi{10.1140/epjst/e2015-02528-2}.

\bibitem[Krishnapriyan et~al.(2021)Krishnapriyan, Gholami, Zhe, Kirby, and
  Mahoney]{NEURIPS2021_df438e52}
Krishnapriyan, A., Gholami, A., Zhe, S., Kirby, R., and Mahoney, M.~W.
\newblock Characterizing possible failure modes in physics-informed neural
  networks.
\newblock In Ranzato, M., Beygelzimer, A., Dauphin, Y., Liang, P., and Vaughan,
  J.~W. (eds.), \emph{Advances in Neural Information Processing Systems},
  volume~34, pp.\  26548--26560. Curran Associates, Inc., 2021.
\newblock URL
  \url{https://proceedings.neurips.cc/paper/2021/file/df438e5206f31600e6ae4af72f2725f1-Paper.pdf}.

\bibitem[Luo \& Clark(2019)Luo and Clark]{luo2019backflow}
Luo, D. and Clark, B.~K.
\newblock Backflow transformations via neural networks for quantum many-body
  wave functions.
\newblock \emph{Physical review letters}, 122\penalty0 (22):\penalty0 226401,
  2019.

\bibitem[Luo et~al.(2021)Luo, Chen, Hu, Zhao, Hur, and Clark]{luo2021gauge}
Luo, D., Chen, Z., Hu, K., Zhao, Z., Hur, V.~M., and Clark, B.~K.
\newblock Gauge invariant autoregressive neural networks for quantum lattice
  models.
\newblock \emph{arXiv preprint arXiv:2101.07243}, 2021.

\bibitem[Luo et~al.(2022{\natexlab{a}})Luo, Chen, Carrasquilla, and
  Clark]{PhysRevLett.128.090501}
Luo, D., Chen, Z., Carrasquilla, J., and Clark, B.~K.
\newblock Autoregressive neural network for simulating open quantum systems via
  a probabilistic formulation.
\newblock \emph{Phys. Rev. Lett.}, 128:\penalty0 090501, Feb
  2022{\natexlab{a}}.
\newblock \doi{10.1103/PhysRevLett.128.090501}.

\bibitem[Luo et~al.(2022{\natexlab{b}})Luo, Chen, Carrasquilla, and
  Clark]{luo2022autoregressive}
Luo, D., Chen, Z., Carrasquilla, J., and Clark, B.~K.
\newblock Autoregressive neural network for simulating open quantum systems via
  a probabilistic formulation.
\newblock \emph{Physical review letters}, 128\penalty0 (9):\penalty0 090501,
  2022{\natexlab{b}}.

\bibitem[Luo et~al.(2022{\natexlab{c}})Luo, Yuan, Stokes, and
  Clark]{luo2022gauge}
Luo, D., Yuan, S., Stokes, J., and Clark, B.~K.
\newblock Gauge equivariant neural networks for 2+ 1d u (1) gauge theory
  simulations in hamiltonian formulation.
\newblock \emph{arXiv preprint arXiv:2211.03198}, 2022{\natexlab{c}}.

\bibitem[Martin et~al.(2016)Martin, Reining, and
  Ceperley]{martin_reining_ceperley_2016}
Martin, R.~M., Reining, L., and Ceperley, D.~M.
\newblock \emph{Interacting Electrons: Theory and Computational Approaches}.
\newblock Cambridge University Press, 2016.
\newblock \doi{10.1017/CBO9781139050807}.

\bibitem[Martyn et~al.(2022)Martyn, Najafi, and Luo]{martyn2022variational}
Martyn, J.~M., Najafi, K., and Luo, D.
\newblock Variational neural-network ansatz for continuum quantum field theory.
\newblock \emph{arXiv preprint arXiv:2212.00782}, 2022.

\bibitem[McMillan(1965)]{mcmillan1965ground}
McMillan, W.~L.
\newblock Ground state of liquid he 4.
\newblock \emph{Physical Review}, 138\penalty0 (2A):\penalty0 A442, 1965.

\bibitem[Mohamed et~al.(2020)Mohamed, Rosca, Figurnov, and
  Mnih]{mohamed2020monte}
Mohamed, S., Rosca, M., Figurnov, M., and Mnih, A.
\newblock Monte carlo gradient estimation in machine learning.
\newblock \emph{The Journal of Machine Learning Research}, 21\penalty0
  (1):\penalty0 5183--5244, 2020.

\bibitem[Nagy \& Savona(2019)Nagy and Savona]{PhysRevLett.122.250501}
Nagy, A. and Savona, V.
\newblock Variational quantum monte carlo method with a neural-network ansatz
  for open quantum systems.
\newblock \emph{Phys. Rev. Lett.}, 122:\penalty0 250501, Jun 2019.
\newblock \doi{10.1103/PhysRevLett.122.250501}.
\newblock URL \url{https://link.aps.org/doi/10.1103/PhysRevLett.122.250501}.

\bibitem[Nocedal \& Wright(1999)Nocedal and Wright]{LGBFS}
Nocedal, J. and Wright, S.~J. (eds.).
\newblock \emph{Large-Scale Quasi-Newton and Partially Separable Optimization},
  pp.\  222--249.
\newblock Springer New York, New York, NY, 1999.
\newblock ISBN 978-0-387-22742-9.
\newblock \doi{10.1007/0-387-22742-3_9}.
\newblock URL \url{https://doi.org/10.1007/0-387-22742-3_9}.

\bibitem[Pfau et~al.(2020)Pfau, Spencer, Matthews, and
  Foulkes]{PhysRevResearch.2.033429}
Pfau, D., Spencer, J.~S., Matthews, A. G. D.~G., and Foulkes, W. M.~C.
\newblock Ab initio solution of the many-electron schr\"odinger equation with
  deep neural networks.
\newblock \emph{Phys. Rev. Res.}, 2:\penalty0 033429, Sep 2020.
\newblock \doi{10.1103/PhysRevResearch.2.033429}.

\bibitem[Rackauckas \& Nie(2017)Rackauckas and
  Nie]{rackauckas2017differentialequations}
Rackauckas, C. and Nie, Q.
\newblock Differentialequations.jl--a performant and feature-rich ecosystem for
  solving differential equations in julia.
\newblock \emph{Journal of Open Research Software}, 5\penalty0 (1):\penalty0
  15, 2017.

\bibitem[Raissi(2018)]{JMLR:v19:18-046}
Raissi, M.
\newblock Deep hidden physics models: Deep learning of nonlinear partial
  differential equations.
\newblock \emph{Journal of Machine Learning Research}, 19\penalty0
  (25):\penalty0 1--24, 2018.
\newblock URL \url{http://jmlr.org/papers/v19/18-046.html}.

\bibitem[Raissi et~al.(2019)Raissi, Perdikaris, and Karniadakis]{RAISSI2019686}
Raissi, M., Perdikaris, P., and Karniadakis, G.
\newblock Physics-informed neural networks: A deep learning framework for
  solving forward and inverse problems involving nonlinear partial differential
  equations.
\newblock \emph{Journal of Computational Physics}, 378:\penalty0 686--707,
  2019.
\newblock ISSN 0021-9991.
\newblock \doi{https://doi.org/10.1016/j.jcp.2018.10.045}.
\newblock URL
  \url{http://www.sciencedirect.com/science/article/pii/S0021999118307125}.

\bibitem[Reh \& G{\"a}rttner(2022)Reh and G{\"a}rttner]{reh2022variational}
Reh, M. and G{\"a}rttner, M.
\newblock Variational monte carlo approach to partial differential equations
  with neural networks.
\newblock \emph{Machine Learning: Science and Technology}, 3\penalty0
  (4):\penalty0 04LT02, 2022.

\bibitem[Reh et~al.(2021)Reh, Schmitt, and G{\"a}rttner]{reh2021time}
Reh, M., Schmitt, M., and G{\"a}rttner, M.
\newblock Time-dependent variational principle for open quantum systems with
  artificial neural networks.
\newblock \emph{Physical Review Letters}, 127\penalty0 (23):\penalty0 230501,
  2021.

\bibitem[Rezende \& Mohamed(2015)Rezende and Mohamed]{rezende2015variational}
Rezende, D. and Mohamed, S.
\newblock Variational inference with normalizing flows.
\newblock In \emph{International conference on machine learning}, pp.\
  1530--1538. PMLR, 2015.

\bibitem[Shankar(2012)]{shankar2012principles}
Shankar, R.
\newblock \emph{Principles of quantum mechanics}.
\newblock Springer Science \& Business Media, 2012.

\bibitem[Sharir et~al.(2020)Sharir, Levine, Wies, Carleo, and
  Shashua]{PhysRevLett.124.020503}
Sharir, O., Levine, Y., Wies, N., Carleo, G., and Shashua, A.
\newblock Deep autoregressive models for the efficient variational simulation
  of many-body quantum systems.
\newblock \emph{Phys. Rev. Lett.}, 124:\penalty0 020503, Jan 2020.
\newblock \doi{10.1103/PhysRevLett.124.020503}.

\bibitem[Sorella(1998)]{sorella1998green}
Sorella, S.
\newblock Green function monte carlo with stochastic reconfiguration.
\newblock \emph{Physical review letters}, 80\penalty0 (20):\penalty0 4558,
  1998.

\bibitem[Sorella(2001)]{sorella2001generalized}
Sorella, S.
\newblock Generalized lanczos algorithm for variational quantum monte carlo.
\newblock \emph{Physical Review B}, 64\penalty0 (2):\penalty0 024512, 2001.

\bibitem[Verstraete et~al.(2009)Verstraete, Wolf, and
  Ignacio~Cirac]{verstraete2009quantum}
Verstraete, F., Wolf, M.~M., and Ignacio~Cirac, J.
\newblock Quantum computation and quantum-state engineering driven by
  dissipation.
\newblock \emph{Nature physics}, 5\penalty0 (9):\penalty0 633--636, 2009.

\bibitem[Vicentini et~al.(2019)Vicentini, Biella, Regnault, and
  Ciuti]{PhysRevLett.122.250503}
Vicentini, F., Biella, A., Regnault, N., and Ciuti, C.
\newblock Variational neural-network ansatz for steady states in open quantum
  systems.
\newblock \emph{Phys. Rev. Lett.}, 122:\penalty0 250503, Jun 2019.
\newblock \doi{10.1103/PhysRevLett.122.250503}.
\newblock URL \url{https://link.aps.org/doi/10.1103/PhysRevLett.122.250503}.

\bibitem[Vicentini et~al.(2022)Vicentini, Hofmann, Szabó, Wu, Roth, Giuliani,
  Pescia, Nys, Vargas-Calderón, Astrakhantsev, and Carleo]{netket3:2022}
Vicentini, F., Hofmann, D., Szabó, A., Wu, D., Roth, C., Giuliani, C., Pescia,
  G., Nys, J., Vargas-Calderón, V., Astrakhantsev, N., and Carleo, G.
\newblock {NetKet 3: Machine Learning Toolbox for Many-Body Quantum Systems}.
\newblock \emph{SciPost Phys. Codebases}, pp.\ ~7, 2022.
\newblock \doi{10.21468/SciPostPhysCodeb.7}.
\newblock URL \url{https://scipost.org/10.21468/SciPostPhysCodeb.7}.

\bibitem[Weinan et~al.(2021)Weinan, Han, and Jentzen]{weinan2021algorithms}
Weinan, E., Han, J., and Jentzen, A.
\newblock Algorithms for solving high dimensional pdes: from nonlinear monte
  carlo to machine learning.
\newblock \emph{Nonlinearity}, 35\penalty0 (1):\penalty0 278, 2021.

\bibitem[Westerhout et~al.(2020)Westerhout, Astrakhantsev, Tikhonov,
  Katsnelson, and Bagrov]{westerhout2020generalization}
Westerhout, T., Astrakhantsev, N., Tikhonov, K.~S., Katsnelson, M.~I., and
  Bagrov, A.~A.
\newblock Generalization properties of neural network approximations to
  frustrated magnet ground states.
\newblock \emph{Nature communications}, 11\penalty0 (1):\penalty0 1593, 2020.

\bibitem[Yoshioka \& Hamazaki(2019)Yoshioka and Hamazaki]{PhysRevB.99.214306}
Yoshioka, N. and Hamazaki, R.
\newblock Constructing neural stationary states for open quantum many-body
  systems.
\newblock \emph{Phys. Rev. B}, 99:\penalty0 214306, Jun 2019.
\newblock \doi{10.1103/PhysRevB.99.214306}.
\newblock URL \url{https://link.aps.org/doi/10.1103/PhysRevB.99.214306}.

\bibitem[{Zhuang} et~al.(2020){Zhuang}, {Tang}, {Ding}, {Tatikonda}, {Dvornek},
  {Papademetris}, and {Duncan}]{adabeleif}
{Zhuang}, J., {Tang}, T., {Ding}, Y., {Tatikonda}, S., {Dvornek}, N.,
  {Papademetris}, X., and {Duncan}, J.~S.
\newblock {AdaBelief Optimizer: Adapting Stepsizes by the Belief in Observed
  Gradients}.
\newblock \emph{arXiv e-prints}, art. arXiv:2010.07468, October 2020.
\newblock \doi{10.48550/arXiv.2010.07468}.

\end{thebibliography}
\bibliographystyle{icml2023}

\newpage
\appendix
\onecolumn

\section{Further Details about the Q Function Formalism}\label{app: Conversions}

\subsection{Quantum Preliminaries}

\label{sec:quantum_prelim}
Here, we provide a brief and intuitive introduction to the theory of bosonic systems. We intentionally simplify most of the definitions and focus on the important concepts to our study. For an in-depth discussion, please refer to~\citep{shankar2012principles}.
 
We will discuss a few important terms that we use throughout the main text.

\paragraph{Hilbert Space.}
In a closed system, i.e., one which is insulated from the environment, each subsystem’s state can be described as a unit vector in a complex vector space $\mathbb{C}^{n}$, a Hilbert space, for some dimension $n$ which we will from now on denote as the \textit{Hilbert space dimension} or the \textit{degrees of freedom}.

\paragraph{Open Quantum Systems} In an open quantum system, interactions with the environment introduce additional uncertainty about the quantum state of the system. To model open systems, we must thus resort to the \textit{density matrix}. The density matrix is an $n\times n$ positive definite unit-trace complex-valued matrix, where $n$ is the Hilbert space dimension of the system. The space $\mathbb{C}^{n\times n}$ of density matrices is sometimes known as the \textit{double Hilbert space}. This space is spanned by the set of outer products $\ket{b_1}\bra{b_2}$ of basis vectors. The density matrix can be thought of as an operator on the Hilbert space.

\paragraph{Braket notation.}
Such notation is used throughout the text to denote quantum states. Quantum states are elements of a complex vector space $V$, equipped with a Hermitian form. In our work we use the standard Hermitian inner product, which in math notation is $(\mathbf{v},\mathbf{w})=\mathbf{v}^\dag \mathbf{w}$. Here $\dag$ denotes the complex conjugate, for any two vectors $\mathbf{v},\mathbf{w} \in V.$  In physics notation, we write $\mathbf{v}$ as $\ket{\mathbf{v}}$ (known as a \emph{ket}) and likewise for $\mathbf{w}$. We also use the notation $\bra{\mathbf{v}}\equiv \mathbf{v}^\dag$, and call this a \emph{bra}. Then, $\mathbf{v}^\dag \mathbf{w}$ can be written $\bra{\mathbf{v}}\hspace{0.1pt}\ket{\mathbf{w}}$, or more concisely as $\braket{\mathbf{v}}{\mathbf{w}}$. Furthermore, $\ket{\mathbf{v}}\bra{\mathbf{w}}$ denotes the outer product of $\mathbf{v}$ and $\mathbf{w}^\dag$.

\paragraph{Particle number Hilbert space, Vacuum states, the Fock space.}
There exists a special Hilbert space known as the particle number Hilbert space or the \textit{Fock space}. This Hilbert space describes a location, such as a potential well, with varying number of particles.  It is spanned by a countably infinite set of orthonormal basis vectors, which we label $\ket{0}, \ket{1}, \ldots$, where $\ket{n}$ represents a system with $n$ particles. To represent a general element of $\ket{0}, \ket{1}, \ldots$, we will use a Roman letter inside the ket or bra. We denote $\ket{0}$ as the \textit{vaccum state} because it represents a system with no particles.

Although a system with varying number of particles can be described by the particle number Hilbert space, there are many other systems that can be similarly described. For example, a particle confined to move in a 1d potential well can be described by this Hilbert space. In this paper, we use the term \textit{site} to refer to any system with a Hilbert space that is the particle number Hilbert space. For multiple sites, the total Hilbert space is the tensor product of each particle's Hilbert space.

\paragraph{Creation, annihilation operators and Coherent state.} The creation operator $a^{\dagger}$ satisfies $a^{\dagger} \ket{n} = \sqrt{n+1} \ket{n+1}$ and the annihilation operator $a$ satisfies $a \ket{n} = \sqrt{n} \ket{n-1}$ with $a\ket{0}=0$. The coherent state $\ket{\alpha}$ with a complex number $\alpha$ is defined as $\ket{\alpha} = e^{\alpha a^{\dagger} - \alpha^{*}a} \ket{0}$, where $e$ should be interpreted as matrix exponential function. A more practical equivalent definition of the coherent state is \begin{equation*}
    \ket{\alpha} = e^{-|\alpha|^2/2}\sum_{n=0}^\infty \frac{\alpha^n}{\sqrt{n!}}\ket{n}.
\end{equation*}

\paragraph{Compute observables.} In quantum mechanics, a density matrix $\rho$ can be expressed as $\rho=\sum_{n,m}\rho_{n,m}\ket{n}\bra{m}$ and an observable $O$ can be expressed as $O=\sum_{n,m} O_{n,m}\ket{n}\bra{m}$, where both $\rho$ and $O$ can be viewed as Hermitian matrices. It follows that the expectation value of the observable $\langle O \rangle = \text{tr}(\rho O) = \sum_{n,m} O_{n,m} \rho_{m,n}$. 

\paragraph{Tensor products.} Suppose we have two Hilbert spaces $\mathcal{H}_1$ and $\mathcal{H}_2$. For every two kets $\ket{a}\in\mathcal{H}_1$ and $\ket{b}\in\mathcal{H}_2$, there exists a ket $\ket{a}\otimes\ket{b}\in \mathcal{H}_1\otimes \mathcal{H}_2$, where $\ket{a}\otimes\ket{b}$ is the tensor product of $\ket{a}$ and $\ket{b}$ and $\mathcal{H}_1\otimes \mathcal{H}_2$ is the tensor product space $\mathcal{H}_1$ and $\mathcal{H}_2$. The inner product for tensor products of Hilbert spaces is defined as $\left(\bra{a}\otimes\bra{b}\right)\left(\ket{c}\otimes\ket{d}\right) = \braket{a}{c}\braket{b}{d}$, where $\bra{a}\otimes\bra{b}$ is the conjugate transpose of $\ket{a}\otimes\ket{b}$. From this it is clear that the basis kets for the new Hilbert space are all pairs of tensor products of basis kets of the two smaller Hilbert spaces. If an operator $O_1$ acts on $\mathcal{H}_1$ and $O_2$ acts on $\mathcal{H}_2$, then $(O_1\otimes O_2)$ acts on $\mathcal{H}_1\otimes \mathcal{H}_2$ according to $(O_1\otimes O_2)\ket{a}\otimes \ket{b} = (O_1\ket{a})\otimes (O_2\ket{b}).$ Finally, note that we often use shorthands such as $O_1$ or $O_2$ to refer to $O_1\otimes 1$ or $1\otimes O_2$, respectively.

\subsection{Q Function to $\rho$}

In this section, we show that for a given $Q(\alpha,\alpha^*),$ the density matrix $\rho$ corresponding to it is given by\begin{equation}\label{eq:QtoRho}
    \bra{m}\rho\ket{n} = \pi\sqrt{ m!n!}\sum_{k=0}^{\min(m,n)} \frac{Q_{m-k,n-k}}{k!},
\end{equation} where \begin{equation*}
    Q_{a,b}(\alpha,\alpha^*) = \frac{1}{a!b!}\left.\frac{\partial^{a+b}}{\partial^{a}\alpha\, \partial^{b}\alpha^*}Q(\alpha,\alpha^*)\right|_{\alpha=\alpha^*=0}.
\end{equation*}

This result generalizes to multi-site Q functions using the tensor product structure, but for simplicity we consider only a single site here.

From expressing $\bra{\alpha}$ and $\ket{\alpha}$ in terms of Harmonic Oscillator eigenstates, we have that \begin{align*}
    Q(\alpha,\alpha^*) &= \frac{1}{\pi}e^{-\alpha\alpha^*}\sum_{m=0}^\infty \sum_{n=0}^\infty \frac{\bra{m}\rho\ket{n}}{\sqrt{m!n!}}\alpha^{*m}\alpha^n\\
    &= \frac{1}{\pi}\left(\sum_{s=0}^\infty \frac{(-1)^s}{s!} (\alpha\alpha^*)^s\right)\sum_{m=0}^\infty \sum_{n=0}^\infty \frac{\bra{m}\rho\ket{n}}{\sqrt{m!n!}}\alpha^{*m}\alpha^n\\
    &= \frac{1}{\pi}\sum_{s=0}^\infty \sum_{m=0}^\infty \sum_{n=0}^\infty \frac{(-1)^s}{s!}\frac{\bra{m}\rho\ket{n}}{\sqrt{m!n!}}\left({\alpha^*}\right)^{m+s}\alpha^{n+s}
\end{align*} To determine $\bra{m}\rho\ket{n}$, we must thus invert this series. However, since we know the correct form, we can simply substitute Equation \ref{eq:QtoRho} into the expression above and show that it correctly gives $Q(\alpha,\alpha^*):$\begin{align*}
    &\frac{1}{\pi}\sum_{s=0}^\infty \sum_{m=0}^\infty \sum_{n=0}^\infty \frac{(-1)^s}{s!}\frac{\bra{m}\rho\ket{n}}{\sqrt{m!n!}}\left({\alpha^*}\right)^{m+s}\alpha^{n+s}\\
    =& \frac{1}{\pi}\sum_{s=0}^\infty \sum_{m=0}^\infty \sum_{n=0}^\infty \frac{(-1)^s}{s!}\frac{\pi\sqrt{ m!n!}\sum_{k=0}^{\min(m,n)} \frac{Q_{m-k,n-k}}{k!}}{\sqrt{m!n!}}\left({\alpha^*}\right)^{m+s}\alpha^{n+s}\\
    =& \sum_{s=0}^\infty \sum_{m=0}^\infty \sum_{n=0}^\infty \sum_{k=0}^{\min(m,n)} \frac{(-1)^s}{s!}\frac{Q_{m-k,n-k}}{k!}\left({\alpha^*}\right)^{m+s}\alpha^{n+s}.
\end{align*} Setting $a=m+s$ and $b=n+s$ gives \begin{align*}
    \sum_{a=0}^\infty \sum_{b=0}^\infty \sum_{s=0}^{\min(a,b)} \sum_{k=0}^{\min(a,b)-s} \frac{(-1)^s}{s!}\frac{Q_{a-k-s,b-k-s}}{k!}\left({\alpha^*}\right)^{a}\alpha^{b}.
\end{align*} Then, setting $d = s+k$ gives \begin{align*}
    &\sum_{a=0}^\infty \sum_{b=0}^\infty \sum_{d=0}^{\min(a,b)}\sum_{s=0}^{d} \frac{(-1)^s}{s!}\frac{Q_{a-d,b-d}}{(d-s)!}\left({\alpha^*}\right)^{a}\alpha^{b}\\
    =& \sum_{a=0}^\infty \sum_{b=0}^\infty \left({\alpha^*}\right)^{a}\alpha^{b} \sum_{d=0}^{\min(a,b)} Q_{a-d,b-d} \sum_{s=0}^{d} (-1)^s {\binom{d}{s}}.
\end{align*} Now, by the Binomial Theorem, $\sum_{s=0}^{d} (-1)^s {\binom{d}{s}} = (1-1)^d = 0^d$, which is 0 unless $d=1$. So, we get \begin{align*}
    &\sum_{a=0}^\infty \sum_{b=0}^\infty Q_{a,b}(\alpha,\alpha^*)\cdot \left({\alpha^*}\right)^{a}\alpha^{b}\\
    =&\sum_{a=0}^\infty \sum_{b=0}^\infty  \frac{\left({\alpha^*}\right)^{a}\alpha^{b}}{a!b!}\frac{\partial^{a+b}}{\partial^{a}\alpha \partial^{b}\alpha^*}Q(\alpha,\alpha^*)\\
    =& Q(\alpha,\alpha^*),
\end{align*} as desired. The last step comes from the Taylor series representation of $Q$, which is only valid if $Q$ is analytic. So as long as $Q$ is analytic, this result holds.

\subsection{Coherent State Identities}
Here we present a few coherent state identities that prove useful in \ref{sec: rho evolution to Q evolution}.
\begin{align*}
    a^\dagger \ket{\alpha} &= a^\dagger e^{-|\alpha|^2/2}\sum_{n=0}^\infty \frac{\alpha^n}{\sqrt{n!}}\ket{n}\\
    &= e^{-|\alpha|^2/2}\sum_{n=0}^\infty \frac{\alpha^n}{\sqrt{n!}}
    \sqrt{n+1}\ket{n+1}\\
    &= e^{-|\alpha|^2/2}\frac{\partial}{\partial\alpha}\sum_{n=0}^\infty \frac{\alpha^{n+1}}{\sqrt{(n+1)!}}\ket{n+1}\\
    &= e^{-|\alpha|^2/2}\frac{\partial}{\partial\alpha}\sum_{n=0}^\infty \frac{\alpha^{n}}{\sqrt{n!}}\ket{n}\\
    &= e^{-|\alpha|^2/2}\frac{\partial}{\partial\alpha}\sum_{n=0}^\infty \frac{\alpha^{n}}{\sqrt{n!}}\ket{n}\\
    &= \frac{\partial}{\partial\alpha}e^{-|\alpha|^2/2}\sum_{n=0}^\infty \frac{\alpha^{n}}{\sqrt{n!}}\ket{n} -  \left(\frac{\partial}{\partial\alpha}e^{-|\alpha|^2/2}\right)\sum_{n=0}^\infty \frac{\alpha^{n}}{\sqrt{n!}}\ket{n}\\
    &= \frac{\partial}{\partial\alpha}e^{-|\alpha|^2/2}\sum_{n=0}^\infty \frac{\alpha^{n}}{\sqrt{n!}}\ket{n} -  \left(\frac{\partial}{\partial\alpha}e^{-|\alpha|^2/2}\right)\sum_{n=0}^\infty \frac{\alpha^{n}}{\sqrt{n!}}\ket{n}\\
    &= \frac{\partial}{\partial\alpha}e^{-|\alpha|^2/2}\sum_{n=0}^\infty \frac{\alpha^{n}}{\sqrt{n!}}\ket{n} +\frac{\alpha^*}{2}e^{-|\alpha|^2/2}\sum_{n=0}^\infty \frac{\alpha^{n}}{\sqrt{n!}}\ket{n}\\
    &= \left(\frac{\alpha^*}{2}+\frac{\partial}{\partial\alpha}\right)e^{-|\alpha|^2/2}\sum_{n=0}^\infty \frac{\alpha^{n}}{\sqrt{n!}}\ket{n}\\
    &= \left(\frac{\alpha^*}{2}+\frac{\partial}{\partial\alpha}\right)\ket{\alpha}.
\end{align*}

Similarly, \begin{align*}
    \bra{\alpha} a = \left(\frac{\alpha}{2}+\frac{\partial}{\partial\alpha^*}\right)\bra{\alpha}.
\end{align*}

Also, 
\begin{equation}
    \frac{\partial}{\partial \alpha^*}\ket{\alpha} = \frac{\partial}{\partial \alpha^*}e^{-|\alpha|^2/2}\sum_{n=0}^\infty \frac{\alpha^n}{\sqrt{n!}}\ket{n}\\
    = -\frac{\alpha}{2}e^{-|\alpha|^2/2}\sum_{n=0}^\infty \frac{\alpha^n}{\sqrt{n!}}\ket{n}\\
    = -\frac{\alpha}{2}\ket{\alpha}
\end{equation} 
and \begin{align*}
    \frac{\partial}{\partial \alpha}\bra{\alpha} = -\frac{\alpha^*}{2}\bra{\alpha}.
\end{align*}

\subsection{$\rho$ Evolution to Q Function Evolution}\label{sec: rho evolution to Q evolution}
In this section, we demonstrate how to convert to a general equation of motion for a density matrix to an equation of motion for the corresponding Q function using the tensor product structure. This result generalizes to multi-site Q functions, but for simplicity we consider only a single site here.

Note that \begin{align*}
    \bra{\alpha}a^\dagger\hat{O}_1\rho\hat{O}_2\ket{\alpha} &= \alpha^*\bra{\alpha}\hat{O}_1\rho\hat{O}_2\ket{\alpha},\\
    \bra{\alpha}\hat{O}_1\rho\hat{O}_2 a\ket{\alpha} &= \alpha\bra{\alpha}\hat{O}_1\rho\hat{O}_2\ket{\alpha}.
\end{align*} Also, \begin{align*}
    \bra{\alpha}a\hat{O}_1\rho\hat{O}_2\ket{\alpha} &= \left[\left(\frac{\alpha}{2}+\frac{\partial}{\partial\alpha^*}\right)\bra{\alpha}\right]\hat{O}_1\rho\hat{O}_2\ket{\alpha}\\
    &= \left(\frac{\alpha}{2}+\frac{\partial}{\partial\alpha^*}\right)\bra{\alpha}\hat{O}_1\rho\hat{O}_2\ket{\alpha}-\bra{\alpha}\hat{O}_1\rho\hat{O}_2\frac{\partial}{\partial\alpha^*}\ket{\alpha}\\
    &= \left(\frac{\alpha}{2}+\frac{\partial}{\partial\alpha^*}\right)\bra{\alpha}\hat{O}_1\rho\hat{O}_2\ket{\alpha}+\frac{\alpha}{2}\bra{\alpha}\hat{O}_1\rho\hat{O}_2\ket{\alpha}\\
    &= \left(\alpha+\frac{\partial}{\partial\alpha^*}\right)\bra{\alpha}\hat{O}_1\rho\hat{O}_2\ket{\alpha},
\end{align*} and \begin{align*}
    \bra{\alpha}\hat{O}_1\rho\hat{O}_2a^\dagger\ket{\alpha} &= \bra{\alpha}\hat{O}_1\rho\hat{O}_2\left(\frac{\alpha^*}{2}+\frac{\partial}{\partial\alpha}\right)\ket{\alpha}\\
    &= \left(\frac{\alpha^*}{2}+\frac{\partial}{\partial\alpha}\right)\bra{\alpha}\hat{O}_1\rho\hat{O}_2\ket{\alpha}-\left[\frac{\partial}{\partial\alpha}\bra{\alpha}\right]\hat{O}_1\rho\hat{O}_2\ket{\alpha}\\
    &= \left(\frac{\alpha^*}{2}+\frac{\partial}{\partial\alpha}\right)\bra{\alpha}\hat{O}_1\rho\hat{O}_2\ket{\alpha}+\frac{\alpha^*}{2}\bra{\alpha}\hat{O}_1\rho\hat{O}_2\ket{\alpha}\\
    &= \left(\alpha^*+\frac{\partial}{\partial\alpha}\right)\bra{\alpha}\hat{O}_1\rho\hat{O}_2\ket{\alpha}.
\end{align*}

With these results, we now have that for an equation of the form \begin{align*}
    \dot\rho = \sum_{j,k,l,m} c_{j,k,l,m}(a^\dagger)^j a^k\rho (a^\dagger)^l a^m,
\end{align*} we can convert the the Q function equation of motion by inserting $\frac{1}{\pi}\bra{\alpha}\;\ket{\alpha}$ to get \begin{align*}
    &\frac{1}{\pi}\bra{\alpha}\dot\rho\ket{\alpha} = \frac{1}{\pi}\sum_{j,k,l,m} c_{j,k,l,m}\bra{\alpha}(a^\dagger)^j a^k\rho (a^\dagger)^l a^m\ket{\alpha}\\
    \implies &\dot Q(\alpha,\alpha^*) = \frac{1}{\pi}\sum_{j,k,l,m} c_{j,k,l,m}(\alpha^*)^j\bra{\alpha} a^k\rho (a^\dagger)^l a^m\ket{\alpha}\\
    \implies &\dot Q(\alpha,\alpha^*) = \frac{1}{\pi}\sum_{j,k,l,m} c_{j,k,l,m}(\alpha^*)^j\left(\alpha+\frac{\partial}{\partial\alpha^*}\right)^k\bra{\alpha} \rho (a^\dagger)^l a^m\ket{\alpha}\\
    \implies &\dot Q(\alpha,\alpha^*) = \frac{1}{\pi}\sum_{j,k,l,m} c_{j,k,l,m}(\alpha^*)^j\left(\alpha+\frac{\partial}{\partial\alpha^*}\right)^k\alpha^m\bra{\alpha} \rho (a^\dagger)^l\ket{\alpha}\\
    \implies &\dot Q(\alpha,\alpha^*) = \frac{1}{\pi}\sum_{j,k,l,m} c_{j,k,l,m}(\alpha^*)^j\left(\alpha+\frac{\partial}{\partial\alpha^*}\right)^k\alpha^m\left(\alpha^*+\frac{\partial}{\partial\alpha}\right)^l \bra{\alpha} \rho \ket{\alpha}\\
    \implies &\dot Q(\alpha,\alpha^*) = \sum_{j,k,l,m} c_{j,k,l,m}(\alpha^*)^j\left(\alpha+\frac{\partial}{\partial\alpha^*}\right)^k\alpha^m\left(\alpha^*+\frac{\partial}{\partial\alpha}\right)^l Q(\alpha,\alpha^*).
\end{align*}

\subsection{Observable calculation with respect to Q function}
In this section, we demonstrate how to efficiently compute observables by sampling from the Q function using the tensor product structure. This result generalizes to multi-site Q functions, but for simplicity we consider only a single site here.

Consider a general observable $\hat{O}.$ Its expected value given a density matrix $\rho$ is \begin{equation}
    \langle\hat{O}\rangle = \Tr(\hat{O}\rho).
\end{equation} Inserting the coherent state resolution of the identity, we get that \begin{align*}
    \Tr(\hat{O}\rho) &= \int \frac{d\alpha d\alpha^*}{\pi}\Tr(\hat{O}\rho\ket{\alpha}\bra{\alpha})\\
    &= \int \frac{d\alpha d\alpha^*}{\pi}\bra{\alpha}\hat{O}\rho\ket{\alpha}.
\end{align*} Depending on the operator, it may be most useful to insert the resolution of the identity elsewhere.

\textbf{Example} The expected value of  $a^m(a^\dagger)^n$ given a density matrix $\rho$ is 
\begin{align*}
    \langle a^m \left(a^\dagger\right)^n\rangle &= \Tr\left(a^m\left(a^\dagger\right)^n\rho\right) \\
    &= \int \frac{d\alpha d\alpha^*}{\pi}\Tr(a^m\ket{\alpha}\bra{\alpha}\left(a^\dagger\right)^n\rho)\\
    &= \int \frac{d\alpha d\alpha^*}{\pi}\bra{\alpha}\left(a^\dagger\right)^n\rho a^m\ket{\alpha}\\
    &= \int \frac{d\alpha d\alpha^*}{\pi}\; \alpha^m (\alpha^*)^n\bra{\alpha}\rho \ket{\alpha}\\
    &= \int d q d p\; (q+ip)^m (q-ip)^n Q(q,p).
\end{align*}\label{eq:obs} If $m\neq n,$ this is not an observable, but could be made an observable by adding its Hermitian conjugate.

\subsection{Choice of observables}\label{app: why obs}
The Liouvillian is chosen because the differential equation evolution is governed by the Liouvillian. In particular, as Eq. 2 shows that $\dot{Q} = LQ$, when the norm of $LQ$ goes to zero, $\dot{Q}$ approaches zero which is the steady state of interest. Hence, the observable Liouvillian signifies how soon the system evolves to steady state. The centroid is chosen as an observable because it is the macroscopic observable that can be directly measured in the experiment. It behaves as the center of the mass of the system, which naturally connects to the classical limit and provides a good intuition and direct visualization on how the system evolves.

\section{Stochastic Euler-KL Method}
Here we derive Equation \ref{eq:KL_grad} for the control-variance gradient of the KL-Divergence. We start with \begin{equation*}
    KL(Q_{\theta}^{t+dt} || Q_{\mathcal{L}}^{t}) = \int Q_{\theta}^{t+dt} \ln \frac{Q_{\theta}^{t+dt}}{Q_{\mathcal{L}}^{t}}.
\end{equation*} Taking the gradient gives \begin{align*}
    \nabla_\theta KL(Q_{\theta}^{t+dt} || Q_{\mathcal{L}}^{t}) &= \nabla_\theta \int Q_{\theta}^{t+dt} \ln \frac{Q_{\theta}^{t+dt}}{Q_{\mathcal{L}}^{t}}\\
    &= \int \left(\nabla_\theta Q_{\theta}^{t+dt}\right) \ln \frac{Q_{\theta}^{t+dt}}{Q_{\mathcal{L}}^{t}} + \int Q_{\theta}^{t+dt} \nabla_\theta \ln \frac{Q_{\theta}^{t+dt}}{Q_{\mathcal{L}}^{t}}\\
    &= \int Q_{\theta}^{t+dt}\left(\nabla_\theta \ln Q_{\theta}^{t+dt}\right) \ln \frac{Q_{\theta}^{t+dt}}{Q_{\mathcal{L}}^{t}} + \int Q_{\theta}^{t+dt} \nabla_\theta \ln Q_{\theta}^{t+dt}\\
    &= \int Q_{\theta}^{t+dt} \left[\ln \frac{Q_{\theta}^{t+dt}}{Q_{\mathcal{L}}^{t}} + 1\right]\nabla_\theta \ln Q_{\theta}^{t+dt}.
\end{align*} 
Now, note that \begin{equation}
    \int Q_{\theta}^{t+dt}(x) \nabla_\theta \ln Q_{\theta}^{t+dt}(x) = \int Q_{\theta}^{t+dt}(x) \frac{\nabla_\theta Q_{\theta}^{t+dt}(x)}{Q_{\theta}^{t+dt}(x)}\\
     = \int \nabla_\theta Q_{\theta}^{t+dt}(x)\\
     = \nabla_\theta \int Q_{\theta}^{t+dt}(x)\\
     = \nabla_\theta 1\\
     = 0.
\end{equation} 
So, letting 
\begin{equation}
    b = \int Q^{t+dt}_{\theta}\ln\frac{Q_{\theta}^{t+dt}(x)}{Q_{\mathcal{L}}^t(x)}\\
    \approx \frac{1}{N} \sum_{x\sim Q^{t+dt}_{\theta}}\ln\frac{Q_{\theta}^{t+dt}(x)}{Q_{\mathcal{L}}^t(x)},
\end{equation} 
we can subtract a control variance to get\begin{align*}
    \nabla_\theta KL(Q_{\theta}^{t+dt} || Q_{\mathcal{L}}^{t}) &= \int Q_{\theta}^{t+dt} \left[\ln \frac{Q_{\theta}^{t+dt}}{Q_{\mathcal{L}}^{t}} + 1\right]\nabla_\theta \ln Q_{\theta}^{t+dt}\\
    &= \int Q_{\theta}^{t+dt} \left[\ln \frac{Q_{\theta}^{t+dt}}{Q_{\mathcal{L}}^{t}} + 1\right]\nabla_\theta \ln Q_{\theta}^{t+dt} - (b+1)\int Q_{\theta}^{t+dt}(x) \nabla_\theta \ln Q_{\theta}^{t+dt}(x)\\
    &= \int Q_{\theta}^{t+dt} \left[\ln \frac{Q_{\theta}^{t+dt}}{Q_{\mathcal{L}}^{t}} -b\right]\nabla_\theta \ln Q_{\theta}^{t+dt}\\
    &= \int Q_{\theta}^{t+dt} \left[\ln \frac{Q_{\theta}^{t+dt}}{Q_{\mathcal{L}}^{t}} -b\right]\nabla_\theta \ln Q_{\theta}^{t+dt}\\
\end{align*} Finally, approximating the integral gives \begin{align*}
    \nabla_\theta KL(Q_{\theta}^{t+dt} || Q_{\mathcal{L}}^{t}) &= \int Q_{\theta}^{t+dt} \left[\ln \frac{Q_{\theta}^{t+dt}}{Q_{\mathcal{L}}^{t}} -b\right]\nabla_\theta \ln Q_{\theta}^{t+dt}\\
    &\approx \frac{1}{N} \sum_{x\sim Q^{t+dt}_{\theta}} \left[\ln \frac{Q_{\theta}^{t+dt}}{Q_{\mathcal{L}}^{t}} -b\right]\nabla_\theta \ln Q_{\theta}^{t+dt},
\end{align*} as desired.

Similar technique on baseline control variance has been used in the context of reinforcement learning~\cite{mohamed2020monte}. It has been shown that it can reduce the variance of the gradient and helpful for the optimization.

\section{Additional Normalizing Flow Implementation Details}\label{app:Flow details}
\subsection{Affine Coupling Flow}

For our Affine Coupling models, we use the following architecture:

For each Affine Coupling layer, we split each input vector into two equal-sized vectors \texttt{v1 = input[:input.shape[0]//2]} and \texttt{v2 = input[input.shape[0]//2:]}. We then compute two neural networks $s = NN_1(\mathbf{W}_1, v_2)$ and $t = NN_2(\mathbf{W}_2, v_2)$. We then return the concatenation of $e^s v_1 + t$ and $v_2.$

The neural networks used in the Affine Coupling layers are fully-connected feed-forward neural networks. They have a set number of hidden layers of a fixed size. The input is fed into a linear layer with output size equal to the hidden layer size (usually 3) and then fed into a GELU nonlinearity \cite{GELU}. Then, this output is concatenated with the original input vector. We refer to the output concatenated with the previous layer as the ``augmented hidden layer." For each subsequent internal layer, we feed the previous augmented hidden layer into a linear layer with output size equal to the hidden layer size. We then feed this output into a GELU and concatenate the previous augmented hidden layer. Finally, for the final linear layer, the output has size equal to the input dimension of the neural network, and we do not apply a GELU or concatenate the previous augmented hidden layer.

The Affine Coupling flow is then constructed as follows:
We use a unit Gaussian centered at the origin as our flow prior. To transform an input vector from the data coordinate system to the coordinate system of the prior, we apply an Affine Coupling layer and then reverse the order of the input vector. We repeat this process a user specified number of times (usually 3). To transform an input vector from the prior coordinate system to the coordinate system of the data, we apply the inverse of the above transformation.

For the Affine Coupling model, the number of inputs and outputs of the neural networks defining the coupling transforms are equal to the dimension of the probability distribution. Increasing the number of sites will increase the number of inputs and outputs of these neural networks. We do not increase the dimensions of the hidden layers of the neural networks. However, at each hidden layer, we concatenate the previous hidden layer to the current one. As a result, the input is concatenated to every hidden layer, so the dimension of each hidden layer effectively increases. As a result, the number of parameters grows as a quadratic function of the number of sites currently. However, it is feasible to replace the concatenations with skip connection, so that the number of parameters in the hidden layers will stay constant as a function of the number of sites.

\subsection{Convex Potential Flows}

For our implementation of the Convex Potential Flows, we closely follow the methods described in \cite{huang2021convex}. Our architecture is described below.

For our Input Convex Neural Network (ICNN), we follow the ICNN architecture given in section 5 of \cite{huang2021convex}. In particular, given an input vector $x$, we use the following procedure to compute the output of the ICNN:

Following \citet{huang2021convex}, let $L$ denote a linear layer, $L^{+}$ denote a linear layer with positive weights, and $s$ denote a softplus. Also, let $a$ denote an ActNorm layer, as defined in \cite{GLOW} and cat denote concatenation. We first compute $h = L(x)$. Then, for each layer in the network, we set \begin{align}
    \tilde{h} &= s(a(L^{+}(h)+L(x)))\\
    h_{aug} &= a(s(L(x)))\\
    h &= \text{cat}\left(\tilde{h}, h_{aug}\right).
\end{align}. The output of the neural network is then \begin{equation}
    \text{out} = ICNN(x) = a(L^{+}(h)+L(x)).
\end{equation}

Now, let $f(x) = s(a_1)\frac{||x||^2}{2} + s(a_2)ICNN(x)$, where the parameters of $f$ are the parameters of $a_1$, $a_2$, and the parameters of $ICNN$. $f$ is an input-convex function.

We use a unit Gaussian centered on the origin as our prior. To transform an input from the data distribution coordinates to the prior coordinates, we apply $\nabla f$. To transform an input $x$ from the prior distribution coordinates to the data coordinates, we use LGBFS \cite{LGBFS} to find the $y$ that minimizes \begin{equation*}
    f(y) - x\cdot y.
\end{equation*}

For the Convex Potential Flow, the number of inputs to the convex neural network is equal to the dimension of the probability distribution, so increasing the number of sites will increase the number of inputs to the convex potential flow. The internal layer sizes all stay constant, so the only change is the input size. Thus, the number of parameters in the hidden layers is constant as a function of the number of sites and only the number of parameters in the first layer is linear in the number of sites.

\section{Additional Experimental Details}\label{app: Experiment details}

\subsection{Dissipative Harmonic Oscillator system parameters}\label{app: Fokker-Planck Parameters}
As mentioned in the main text, we uniformly sample the dissipative harmonic oscillator system's parameters $\bar{n}_j\in [3,7),$ $\gamma_j\in [0.5,1.5),$ and $\omega_j\in [0.5,1.5).$ 

The values sampled are as follows:

For the 1-site system, the sampled parameters were\begin{lstlisting}
    n_bar = [4.84872804]
    gamma = [1.39682866]
    omega_0 = [1.08564521],
\end{lstlisting} for the 2-site system, the sampled parameters were \begin{lstlisting}
    n_bar = [6.32870528, 5.24684123]
    gamma = [0.85629208, 1.14026682]
    omega_0 = [1.04436318, 0.75820899],
\end{lstlisting} and for the 20-site system, the sampled parameters were \begin{lstlisting}
    n_bar = [5.56599508, 3.65962215, 4.00906784, 4.18744726, 
             6.60341254, 6.11329932, 4.84059527, 5.81464032, 
             5.79018256, 3.86014355, 6.84045506, 4.05790151, 
             6.92639748, 3.43788247, 4.17439805, 5.9303111,  
             5.63412769, 4.52153322, 3.56601688, 4.38005014]
             
    gamma = [0.71584443, 1.08686172, 1.40588976, 1.47121715,
             0.87775305, 0.78137437, 1.27848082, 1.14247345,
             0.95718403, 0.76484186, 1.22056516, 1.24775589,
             0.57332893, 1.06557609, 0.60105471, 1.32710909, 
             0.90712674, 0.67560123, 0.98142727, 0.84515189]
    
    omega_0 = [1.48982032, 0.52079238, 1.30285575, 0.89810077,
               1.35623683, 0.78528379, 1.0019163,  0.77954035,
               0.93705822, 1.32502792, 0.53982753, 0.67051701,
               0.95749435, 0.98833336, 0.86078757, 1.00879361,
               0.88153798, 1.05195061, 1.17483548, 1.1718404].
\end{lstlisting}

\subsection{Pseudo-spectral and finite difference baseline details}
As a baseline approach to solving the Q function evolution PDE (Eq.~\ref{eq:openQ}), we implement a pseudo-spectral and finite difference discretization of the PDE \citep{fornberg1998practical} in a square domain with $-10 < q_j < 10$ and $-10 < p_j < 10$ for each site $j$, set $Q=0$ at the boundaries, and integrate using an adaptive Tsitouras 5/4 Runge-Kutta solver (Tsit5) \cite{rackauckas2017differentialequations} while projecting at each time step to ensure the probability density $Q$ remains positive and normalized. The psuedo-spectral method uses periodic boundary conditions and computes spatial derivatives using a fast Fourier transform. The finite difference method uses Dirichlet boundary conditions set at zero and computes spatial derivatives using the standard second-order finite difference stencil. We use a grid size of $256$ grid points per dimension for 1-site and $32$ grid points per dimension for 2-sites, resulting in a state size of $256^2 = \text{65,536}$ for 1-site and $32^4 = \text{1,048,576}$ for 2-sites. Note that for a fixed grid size, the state grows exponentially with the number of sites---i.e.,\ the curse of dimensionality. This limits our ability to perform more fine-grained simulations on larger domains and makes this baseline approach intractable for more than a few sites.

\subsection{PINN baseline details}

We also use Physics Informed Neural Networks (PINNs) as a baseline. To implement this, we use the \hyperlink{https://github.com/mathLab/PINA}{PINA} library which is built on top of PyTorch. For each problem, we have two loss terms. The first computes the $L_2$ loss between the predicted initial distribution and the actual initial distribution for points sampled uniformly from within the domain of the solver at $t=0$. The second computes the $L_2$ loss between the PINN time derivative and the predicted time derivative $\tilde{\mathcal{L}}Q$ at points sampled uniformly from within the spacial and temporal domain of the solver. The total loss is the sum of these two losses. We then optimize using gradient decent. Every 500 epochs we re-sample the points with which to compute the loss. 

For each experiment, we use a fully connected feed forward neural network. The network has layer sizes of [input size, 40, 40, 40, 1]. For the first layer, we feed the input through a linear layer with output dimension 40 and then apply a GELU nonlinearity. We then concatenate the input. We denote a hidden layer with the previous layer concatenated the ``augmented hidden layer." For each subsequent layer, we take the input, feed it through a linear layer with output the size of the next hidden layer, apply GELU, and then concatenate the previous augmented linear layer. For the final layer, we apply a linear layer with output dimension 1 and do not apply GELU or concatenate.

The following are the hyperparameters used for each of the experiments:

\begin{itemize}
    \item \textbf{1-site Harmonic Oscillator:} We use 1000 samples at a time for the initial condition and 50000 samples at a time for the derivative condition. We train for 25000 epochs with a learning rate of 0.001.

    \item \textbf{2-site Harmonic Oscillator:} We use 1000 samples at a time for the initial condition and 30000 samples at a time for the derivative condition. We train for 25000 epochs with a learning rate of 0.001.

    \item \textbf{20-site Harmonic Oscillator:} We use 3000 samples at a time for the initial condition and 3000 samples at a time for the derivative condition. We train for 50000 epochs with a learning rate of 0.001. Here, we have to decrease the number of samples for the derivative condition because of memory limits.

    \item \textbf{2-site Dissipative Bosonic Model:} We use 1000 samples at a time for the initial condition and 30000 samples at a time for the derivative condition. We train for 25000 epochs with a learning rate of 0.001.
\end{itemize}

\subsection{Stochastic baseline details}

Using the known Green's function for the $N$-site Harmonic Oscillator system \cite{QuantumOpticsChapter4}, we can construct a stochastic differential equation (SDE)
\begin{equation}
\begin{aligned}
    dq_i &= (-\gamma_i q_i/2 + \omega_0 p_i)\,dt + \sqrt{\gamma_i(\bar n_i+1)/2}\,dW_{q_i}\\
    dp_i &= (-\gamma_i p_i/2 - \omega_0 q_i)\,dt + \sqrt{\gamma_i(\bar n_i+1)/2}\,dW_{p_i},
\end{aligned}
\end{equation}
where $dW_{q_i}, dW_{p_i}$ are independent Wiener processes with unit variance and $i\in \{1,\ldots,N\}$. Starting with samples from the initial Q function, this SDE generates samples $q_i, p_i$ from $Q$ at each time point, which can be used to compute simple observables, such as the centroid (i.e.,\ the mean of the samples). Note that this approach does not explicitly provide the Q function and so cannot be used to compute observables involving $Q$ or derivatives of $Q$. This method also only works for a limited set of systems whose evolution equations admit a stochastic description, e.g.,\ a Fokker--Planck equation.

In our experiments, we use 100,000 sample points for the 1-site and 2-site Harmonic Oscillators and 10,000 sample points for the 20-site Harmonic Oscillator.

\subsection{Flow Initialization details}

As discussed in Section \ref{sec: BH} of the main text, for the Bose Hubbard simulation, we initialize our Normalizing Flow models to the desired initial state of \begin{equation*}
    Q = \frac{\left[(q_1-q_2)^2+(p_1-p_2)^2\right]^{100}}{\pi^2\cdot 2^{100}\cdot 100!} e^{-(q_1^2+p_1^2+q_2^2+p_2^2)}.
\end{equation*} To do this, we use the two pretraining methods described in Section \ref{sec: pretraining}. We describe our pretraining hyperparameters in more detail below. At any point below, if we mention sampling from the exact distribution, we do so using MCMC algorithms.

We first initialize the Convex Potential Flow ActNorm layers by providing 10000 samples from the exact distribution. Next, we perform 200 epochs of the following training algorithm (we use a learning rate of 1e-2):

\begin{enumerate}
    \item Sample 1000 points from the exact distribution $Q_{\text{init}}$.
    \item Compute the loss $L = -\sum_{x\sim Q_{\text{init}}}\ln Q_\theta (x).$
    \item Backpropagate to obtain $\nabla_\theta L.$
    \item Take a gradient step using the Adabeleif optimizer \cite{adabeleif}.
\end{enumerate}

After this, we perform 3000 training steps of the following training algorithm (we use a learning rate of 1e-3):

\begin{enumerate}
    \item Sample 1000 points from the model distribution $Q_\theta$.
    \item Compute the gradient update \begin{equation}
        \nabla_\theta KL \approx -\frac{1}{N} \sum_{x\sim Q_{\theta}}\frac{Q_{\text{init}}(x)}{Q_{\theta}(x)}\nabla_\theta\ln Q_{\theta}(x).
    \end{equation}
    \item Take a gradient step using the Adabeleif optimizer \cite{adabeleif}.
\end{enumerate}

Note that for the dissipative harmonic oscillator systems, we use the exact initial state as the prior and initialize the normalizing flow transformation to be the identity. As such, we do not need to use pretraining for the dissipative harmonic oscillator systems.

\subsection{Euler experiment details}

Below are the hyperparameters we use for the Euler method. For the Harmonic Oscillator results, we use a 3 layer RealNVP where each affine transformation is a 2-hidden-layer feed-forward neural network with hidden layers of size 5. Instead of skip connections in the feed-forward neural network, we concatenate the previous activations at each activation layer. For the Dissipative Bosonic Model result, we use a Convex Potential Flow with a 5-hidden-layer input-convex neural network with hidden layers of size 20 and augmented layers of size 4, see \cite{huang2021convex}.

\begin{itemize}
    \item \textbf{1-site Harmonic Oscillator:} We train for 1500 steps with a step size of 0.01. For each step, we use the KL control variance loss to fit for 150 epochs with a learning rate of 0.001. We use 1000 samples per fitting epoch.

    \item \textbf{2-site Harmonic Oscillator:} We train for 1500 steps with a step size of 0.01. For each step, we use the KL control variance loss to fit for 150 epochs with a learning rate of 0.001. We use 1000 samples per fitting epoch.

    \item \textbf{20-site Harmonic Oscillator:} We train for 1500 steps with a step size of 0.01. For each step, we use the KL control variance loss to fit for 150 epochs with a learning rate of 0.001. We use 10000 samples per fitting epoch.

    \item \textbf{2-site Dissipative Bosonic Model:} We train for 400 steps with a step size of 0.02. For each step, we use the KL control variance loss to fit for 200 epochs with a learning rate of 0.002. We use 10000 samples per fitting epoch.
\end{itemize}

\subsection{TDVP experiment details}

Below are the hyperparameters we use for the TDVP method. For the Harmonic Oscillator results, we use a 3 layer RealNVP where each affine transformation is a 2-hidden-layer feed-forward neural network with hidden layers of size 5. Instead of skip connections in the feed-forward neural network, we concatenate the previous activations at each activation layer. For the Dissipative Bosonic Model result, we use a Convex Potential Flow with a 5-hidden-layer input-convex neural network with hidden layers of size 20 and augmented layers of size 4, see \cite{huang2021convex}.

\begin{itemize}
    \item \textbf{1-site Harmonic Oscillator:} We train for 1500 steps with a step size of 0.01. We use 1000 samples per step. We use a diagonal shift of 0.01.

    \item \textbf{2-site Harmonic Oscillator:} We train for 1500 steps with a step size of 0.01. We use 1000 samples per step. We use a diagonal shift of 0.01.

    \item \textbf{20-site Harmonic Oscillator:} We train for 1500 steps with a step size of 0.01. We use 10000 samples per step. We use a diagonal shift of 0.01.

    \item \textbf{2-site Dissipative Bosonic Model:} We train for 2000 steps with a step size of 0.004. We use 10000 samples per step. We use a diagonal shift of 0.01.
\end{itemize}

\section{Additional Experimental Results}\label{app:additional experiment}

\begin{figure*}[t]
    \centering
    
    \includegraphics[width=\linewidth]{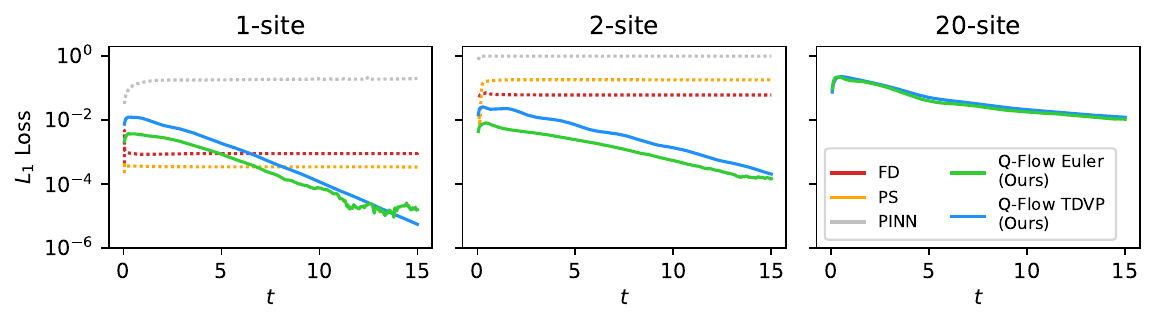}
    
    \caption{The L1 divergence between the simulated Q function and the exact Q function for 1-site, 2-site, and 20-site dissipative harmonic oscillators. Error bars are included for all but the finite difference and pseudo-spectral results but are small for most observables. In the 20-site case, it is not possible to run finite difference (FD) or pseudo-spectral methods (PS). Further, although the PINN method runs, it produces an $L_1$ loss on the order of $10^{30}$, so we do not display it here.}
    
    \label{fig: Fokker-Planck Divergence}
\end{figure*}

Figure \ref{fig: Fokker-Planck Divergence} displays the $L_1$ divergence between the simulated Q function and the exact dissipative harmonic oscillator Q function for various simulation methods. Table \ref{table: Fokker-Planck L1 Loss Appendix} shows the same information as \ref{table: Fokker-Planck L1 Loss} but with errors included. These give additional information about the $L_1$ Loss evolution for the dissipative harmonic oscillator.

\begin{table}[t]
\centering

\begin{tabular}{cccccc}

\toprule
\multicolumn{6}{c}{1-site}\\
\cmidrule(r){1-6}
& Q-Flow  & Q-Flow & & & \\
Time & Euler (ours) & TDVP (ours) & PINN & PS & FD \\
\cmidrule(r){1-1}
\cmidrule(r){2-2}
\cmidrule(r){3-3}
\cmidrule(r){4-4}
\cmidrule(r){5-5}
\cmidrule(r){6-6}

3 & $(2.08\pm 0.01)\cdot 10^{-3}$ & $(5.11\pm 0.01)\cdot 10^{-3}$ & $(1.79\pm 0.01)\cdot 10^{-1}$ & $\mathbf{3.47\cdot 10^{-4}}$ & $8.90\cdot 10^{-4}$\\ 
6 & $(5.10\pm 0.01)\cdot 10^{-4}$ & $(1.17\pm 0.00)\cdot 10^{-3}$ & $(1.84\pm 0.01)\cdot 10^{-1}$ & $\mathbf{3.47\cdot 10^{-4}}$ & $9.01\cdot 10^{-4}$\\ 
9 & $\mathbf{(1.01\pm 0.00)\cdot 10^{-4}}$ & $(2.16\pm 0.01)\cdot 10^{-4}$ & $(1.91\pm 0.01)\cdot 10^{-1}$ & $3.47\cdot 10^{-4}$ & $9.01\cdot 10^{-4}$\\ 
12 & $\mathbf{(1.68\pm 0.01)\cdot 10^{-5}}$ & $(3.58\pm 0.01)\cdot 10^{-5}$ & $(1.91\pm 0.01)\cdot 10^{-1}$ & $3.47\cdot 10^{-4}$ & $9.01\cdot 10^{-4}$\\ 
15 & $(1.58\pm 0.01)\cdot 10^{-5}$ & $\mathbf{(5.55\pm 0.01)\cdot 10^{-6}}$ & $(1.98\pm 0.01)\cdot 10^{-1}$ & $3.47\cdot 10^{-4}$ & $9.01\cdot 10^{-4}$\\ 

\cmidrule(r){1-6}
\multicolumn{6}{c}{2-site}\\
\cmidrule(r){1-6}

3 & $\mathbf{(3.91\pm 0.01)\cdot 10^{-3}}$ & $(1.23\pm 0.00)\cdot 10^{-2}$ & $1.00\pm 0.00$ & $1.83\cdot 10^{-1}$ & $6.12\cdot 10^{-2}$\\ 
6 & $\mathbf{(1.91\pm 0.00)\cdot 10^{-3}}$ & $(4.66\pm 0.01)\cdot 10^{-3}$ & $1.00\pm 0.00$ & $1.82\cdot 10^{-1}$ & $6.09\cdot 10^{-2}$\\ 
9 & $\mathbf{(7.59\pm 0.02)\cdot 10^{-4}}$ & $(1.77\pm 0.00)\cdot 10^{-3}$ & $1.00\pm 0.00$ &  $1.81\cdot 10^{-1}$ & $6.09\cdot 10^{-2}$\\ 
12 & $\mathbf{(2.92\pm 0.01)\cdot 10^{-4}}$ & $(6.21\pm 0.01)\cdot 10^{-4}$ & $1.00\pm 0.00$ &  $1.81\cdot 10^{-1}$ & $6.09\cdot 10^{-2}$\\ 
15 & $\mathbf{(1.47\pm 0.00)\cdot 10^{-4}}$ & $(2.05\pm 0.00)\cdot 10^{-4}$ & $1.00\pm 0.00$ &  $1.81\cdot 10^{-1}$ & $6.09\cdot 10^{-2}$\\ 

\cmidrule(r){1-6}
\multicolumn{6}{c}{20-site}\\
\cmidrule(r){1-6}

3 & $\mathbf{(9.94\pm 0.03)\cdot 10^{-2}}$ & $(1.08\pm 0.00)\cdot 10^{-1}$ & $(2.17\pm 0.62)\cdot 10^{31}$ &  - & -\\ 
6 & $\mathbf{(3.29\pm 0.01)\cdot 10^{-2}}$ & $(4.10\pm 0.01)\cdot 10^{-2}$ & $(2.38\pm 0.81)\cdot 10^{30}$ &  - & -\\ 
9 & $\mathbf{(2.02\pm 0.01)\cdot 10^{-2}}$ & $(2.44\pm 0.01)\cdot 10^{-2}$ & $(1.34\pm 0.52)\cdot 10^{29}$ &  - & -\\ 
12 & $\mathbf{(1.46\pm 0.00)\cdot 10^{-2}}$ & $(1.68\pm 0.00)\cdot 10^{-2}$ & $(1.46\pm 1.19)\cdot 10^{28}$ &  - & -\\ 
15 & $\mathbf{(1.07\pm 0.00)\cdot 10^{-2}}$ & $(1.23\pm 0.00)\cdot 10^{-2}$ & $(7.07\pm 3.66)\cdot 10^{26}$ &  - & -\\ 

\bottomrule
\end{tabular}

\caption{$L_1[Q_{\text{sim}},Q_{\text{exact}}]$ for each simulation method over time, with errors. For each row, we mark the best result in bold.}
\label{table: Fokker-Planck L1 Loss Appendix}
\end{table}

\end{document}